\documentclass[preprint,1p,12pt]{elsarticle}

\usepackage[ruled,vlined,linesnumbered]{algorithm2e}

\usepackage{nicefrac,hyperref}
\usepackage{amssymb,amsmath, stmaryrd}
\usepackage[latin1]{inputenc}
\usepackage{color}
\usepackage{pict2e,picture}
\usepackage [ all ]{xy}
\usepackage{epsfig}
\usepackage{xcolor,tikz}
\usetikzlibrary{fit}
\usepackage{verbatim}
\usepackage[normalem]{ulem}
\usepackage{soul}
\newcommand{\ct}[1]{\cred{\textst{#1}}}
\definecolor{naranjauca}{cmyk}{ 0, 0.6, 1, 0}

\newcommand{\refig}[1]{Figure~\ref{#1}}

\newcommand{\adjoint}{\mathop{\&}\nolimits}

\newcommand{\cred}[1]{{\color{red} #1}}

\newcommand{\tss}{\hbox{small size}}
\newcommand{\tms}{\hbox{medium size}}
\newcommand{\tls}{\hbox{large size}}
\newcommand{\tns}{\hbox{near sun}}
\newcommand{\tfs}{\hbox{far sun}}
\newcommand{\tmy}{\hbox{moon yes}}
\newcommand{\tmn}{\hbox{moon no}}

\newcommand{\mss}{\text{ss}}
\newcommand{\ms}{\text{ms}}
\newcommand{\ls}{\text{ls}}
\newcommand{\ns}{\text{ns}}
\newcommand{\fs}{\text{fs}}
\newcommand{\my}{\text{my}}
\newcommand{\mn}{\text{mn}}

\newtheorem{theorem}{Theorem}
\newtheorem{corollary}[theorem]{Corollary}
\newtheorem{lemma}[theorem]{Lemma}
\newtheorem{proposition}[theorem]{Proposition}
\newdefinition{definition}[theorem]{Definition }
\newdefinition{remark}[theorem]{Remark }
\newdefinition{example}[theorem]{Example }
\newproof{proof}{Proof}

\newcommand{\figlat}[2][0.8]{
	\tikzstyle{whitenode}=[circle,draw=black!75,fill=white!20, text width= 9pt]
	\tikzstyle{blacknode}=[circle,draw=black!75,fill=black!75]
	\begin{tikzpicture}[inner sep=0.75mm,scale=#1, every node/.style={scale=0.75}]		
	\node at (0,0) (C0) [whitenode] {$C_0$};  			
	\node at (0,1) (C2) [whitenode] {$C_2$};
	\node at (1,1) (C4) [whitenode] {$C_4$};
	\node at (-0.9,1) (C1) [whitenode] {$C_1$};
	\node at (-2,1) (C5) [whitenode] {$C_5$};
	\node at (2,1) (C3) [whitenode] {$C_3$};
	
	\draw [-] (C1) -- (C0);
	\draw [-] (C2) -- (C0);
	\draw [-] (C3) -- (C0);
	\draw [-] (C4) -- (C0);
	\draw [-] (C5) -- (C0);
	
	\node at (0,2) (C6) [whitenode] {$C_6$};  			
	\node at (1,2) (C11) [whitenode, text width=12pt] {$C_{11}$};
	\node at (-1,2) (C7) [whitenode] {$C_7$};
	
	\draw [-] (C7) -- (C5);
	\draw [-] (C7) -- (C1);
	\draw [-] (C6) -- (C1);
	\draw [-] (C6) -- (C2);
	\draw [-] (C11) -- (C2);		
	\draw [-] (C11) -- (C4);
	\draw [-] (C11) -- (C3);
	
	\node at (1,3) (C8) [whitenode] {$C_8$};
	\node at (-1,3) (C9) [whitenode] {$C_9$};
	\node at (0,4) (C10) [whitenode,text width=12pt] {$C_{10}$};
	\draw [-] (C9) -- (C7);
	\draw [-] (C9) -- (C6);
	\draw [-] (C8) -- (C6);
	\draw [-] (C8) -- (C11);
	\draw [-] (C10) -- (C9);
	\draw [-] (C10) -- (C8);
	{#2}
	\end{tikzpicture}
}
\newcommand{\figlatnodes}[2][0.75]{
	\tikzstyle{place}=[circle,draw=black!75,fill=black!75]
	\tikzstyle{transition}=[circle,draw=black!75,fill=black!75]
	
	\begin{tikzpicture}[inner sep=0.5mm,scale=#1, every node/.style={scale=#1}]		
	\node at (0,0) (C0) [place, label=270:$\bot$] {};  			
	\node at (0,1) (C2) [transition, label=left:$c$] {};
	\node at (1,1) (C4) [transition, label=right:$d$] {};
	\node at (-1,1) (C1) [transition,label=left:$b$] {};
	\node at (-2,1) (C5) [transition, label=left:$a$] {};
	\node at (2,1) (C3) [transition, label=right:$e$] {};
	
	\draw [-] (C1) -- (C0);
	\draw [-] (C2) -- (C0);
	\draw [-] (C3) -- (C0);
	\draw [-] (C4) -- (C0);
	\draw [-] (C5) -- (C0);
	
	\node at (0,2) (C6) [transition, label=above:$g$] {};  			
	\node at (1,2) (C11) [transition, label=right:$h$] {};
	\node at (-1,2) (C7) [transition, label=left:$f$] {};
	
	\draw [-] (C7) -- (C5);
	\draw [-] (C7) -- (C1);
	\draw [-] (C6) -- (C1);
	\draw [-] (C6) -- (C2);
	\draw [-] (C11) -- (C2);
	\draw [-] (C11) -- (C4);
	\draw [-] (C11) -- (C3);
	
	\node at (1,3) (C8) [transition,label=right:$j$] {};
	\node at (-1,3) (C9) [transition, label=left:$i$] {};
	\node at (0,4) (C10) [transition, label=above:$\top$] {};
	\draw [-] (C9) -- (C7);
	\draw [-] (C9) -- (C6);
	\draw [-] (C8) -- (C6);
	\draw [-] (C8) -- (C11);
	\draw [-] (C10) -- (C9);
	\draw [-] (C10) -- (C8);
	{#2}
	\end{tikzpicture}
}
\newcommand{\minipagethree}[3]{
	\begin{minipage}{0.32\textwidth}
		\begin{center}
			#1
		\end{center}
	\end{minipage}
	\begin{minipage}{0.32\textwidth}
		\begin{center}
			#2
		\end{center}
	\end{minipage}
	\begin{minipage}{0.32\textwidth}
		\begin{center}
			#3
		\end{center}
	\end{minipage}
}
\newcommand{\minipagetwo}[2]{
	\begin{minipage}{0.45\textwidth}
		\begin{center}
			#1
		\end{center}
	\end{minipage}
	\begin{minipage}{0.45\textwidth}
		\begin{center}
			#2
		\end{center}
	\end{minipage}
}

\journal{Fuzzy Sets and Systems}

\makeatletter
\def\ps@pprintTitle{%
  \let\@oddhead\@empty
  \let\@evenhead\@empty
  \let\@evenfoot\@oddfoot
}
\makeatother

\begin{document}

	\begin{frontmatter}
		
		\title{Reducing concept lattices by means of\\ a weaker notion of congruence}

		\author
		{Roberto G. Arag\'on, Jes\'{u}s Medina, Elo\'isa Ram\'irez-Poussa}

		\address
		{Department of Mathematics,
			University of  C\'adiz. Spain\\
			Email: \texttt{\{roberto.aragon,jesus.medina,eloisa.ramirez\}@uca.es}}

	\begin{abstract}
	
	Attribute and size reductions are  key issues in formal concept analysis. In this paper, we consider a special kind of equivalence relation to reduce concept lattices, which will be called  local congruence. This equivalence relation is based on the notion of congruence on lattices, with the goal of losing as less information as possible  and being suitable for the reduction of concept lattices. We analyze how the equivalence classes obtained from a local congruence can be ordered. Moreover, different  properties related to the algebraic structure of the whole set of local congruences are also presented. Finally, a procedure to reduce    concept lattices by the new weaker notion of congruence is introduced. This procedure can be applied to the classical and fuzzy  formal concept analysis frameworks.
		
\end{abstract}

	\end{frontmatter}

	\section{Introduction}

Formal Concept Analysis (FCA) is an exploratory data analysis technique, introduced by Ganter and Wille in~\cite{GanterW}, which has been widely studied from  theoretical and applied perspectives. One of the key problems of formal concept analysis is  to reduce the computational complexity of computing the complete lattice associated with the considered formal context (dataset).  One procedure to address this problem is to find mechanisms to reduce the number of attributes, preserving the most important information contained in the context. Indeed,  we can find many works which analyze different mechanisms that chase this goal~\cite{buruscoipmu2018,Antoni2016,chen2019,ar:ins:2015,Cornejo2017,ins2018:cmr,konecny2019,camwa-medina,Ren2016,Shao2017}. 

Recently, in~\cite{bmrd:RSTFCA:c,bmrd:FCARST:f}, the authors have presented   novel mechanisms to reduce classical and fuzzy formal contexts
based on the reduction philosophy considered in Rough Set Theory, which is another mathematical theory closely related to FCA~\cite{ins-medina,camwa-medina}. In the aforementioned papers, the authors exposed that when the number of attributes of a context
is reduced, an equivalence relation on the set of concepts of the original concept lattice is induced, both in the classical and fuzzy cases. In addition, they also showed that the resulting equivalence classes have the structure of join-semilattices with a maximum element. In the light of the results presented in~\cite{bmrd:RSTFCA:c,bmrd:FCARST:f}, it is natural to ask  how we could complement  the introduced reduction mechanisms in order to ensure that the obtained equivalence classes  are closed algebraic structures. 

Specifically, we are interested in obtaining equivalence classes satisfying that they  are convex sublattices of the original concept lattice. This target can be reached by considering the notion of congruence relation on lattices~\cite{Blyth, DaveyPriestley,GratzerGLT,GratzerUA}. Although  congruence relations within the environment of FCA have not been studied extensively, we can find some works that analyze the use of congruence relations within this mathematical theory.  For example, congruence relations have been applied in lattice/context decomposition as Atlas decomposition~\cite{GanterW}, the subdirect decomposition~\cite{viaud145} or the reverse doubling construction~\cite{Viaud2015}. In addition, the  links between implications and congruence relations have been analyzed in~\cite{VIAUD} and congruence relation have proved to be suitable to handle with inconsistent formal decision contexts~\cite{LiWang2017}. 

However, a significant amount of information can be lost when congruence relations are considered to reduce concept lattices, due mainly to the  restrictions imposed by the quadrilateral-closed property. In order to address this issue, in this paper we continue with the study presented in~\cite{aragonESCIM19}, in which we introduced a weaker notion of congruence by means of  the elimination of the {aforementioned restrictive property}.
We  {will} analyze how the equivalence classes obtained from a local congruence can be ordered {and so, if some hierarchy exists among the clusters provided by the local congruence. Then, since  different local congruence relations  can be defined on a concept lattice, we will go further to a meta level, studying the algebraic structure of the set of all local congruences that can be defined on a lattice and other interesting properties.}
Finally, based on the obtained results, a procedure to reduce concept lattices is presented by using local congruence relations. One of the advantages provided by this procedure is that it can be applied both in the classical and fuzzy generalizations of  formal  concept analysis. In this work, examples to illustrate the proposed procedure are also included. The introduced examples  consider classical FCA, as well as the fuzzy generalization of this theory provided by the multi-adjoint framework~\cite{mor-fss-cmpi}. 
 These examples also confirm that  the use of this kind of equivalence relations is more suitable for this task than the use of congruences, since the amount of lost information  is minimized.

The paper is organized as follows:  Section~\ref{preliminares} reviews some preliminary notions related to lattice theory, congruences on lattices and formal concept analysis, which are necessary to follow this work. In Section~\ref{def:wc:red} the notion of local congruence {and several properties are introduced}. {A study on the ordering among the equivalence classes obtained from local congruences is included in Section~\ref{sec:poset}. The properties related to the algebraic structure of the whole set of local congruences that can be defined on a lattice and to principal local congruences are given in Section~\ref{algest:wc}.}  {Section~\ref{sec:reducmec}  presents} a mechanism to reduce concept lattices based on the use of local congruences. The paper ends in Section~\ref{conclusiones}, showing some conclusions and proposing {diverse} future {challenges.}

	\section{Preliminaries}\label{preliminares}
	
	In this section, some preliminary notions used in this work will be recalled {and we will state the considered notation.} 
	
	{We will consider a lattice as an algebraic structure $(L, \wedge,\vee)$ and as an ordered set $(L, \preceq)$. It is well known that these two points of view are equivalent, since we can define the two operators infimum and supremum from the partial order and vice versa, see The Connecting Lemma in \cite{DaveyPriestley}. Therefore, we will write $(L, \wedge,\vee)$ or $(L, \preceq)$ indistinctly, depending on the most suitable point of view in each case.}

	 In this work, we are interested in defining equivalence  relations  on complete lattices.
	 We will write $(a,b)\in R$ with $a, b\in A$ to indicate that $a$ and $b$ are related under the  {binary} relation $R$. 
	Notice that an equivalence relation $R$ on $A$ gives rise to a partition of $A$, whose subsets are the equivalence classes of $R$. The set of all the equivalence classes of $R$ is called \emph{quotient set} and it is denoted as $A/R$. Equivalently, a partition of $A$ gives rise to an equivalence relation whose equivalence classes are the subsets of the partition.
	
	From this point forward if $\rho \subseteq A\times A$ is an equivalence relation on a set $A$, we will denote the equivalence class of an element $a\in A$ as $[a]_\rho = \{b\in A \mid (a,b)\in \rho\}$.

\subsection{ Congruence on lattices}

This section introduces the notion of congruence on a lattice and some features which are essential to develop our work. First of all, we present the definition of equivalence relation that is compatible with the operation of the algebraic structure.

\begin{definition}\label{compatibilidad}
		 We say that an equivalence relation $\theta$ on a given {lattice $(L, \wedge, \vee)$} is \emph{compatible} with the {supremum $\vee$ and the infimum $\wedge$ of the lattice} if, for all $a, b, c, d\in L$,
		$$(a,b)\in \theta ~\text{ and }~(c,d)\in \theta$$
		imply 
		$$(a\vee c, b \vee d) \in \theta  ~\text{ and }~ (a\wedge c,  b \wedge d) \in \theta$$ 
\end{definition}

We can now state the definition of congruence on a lattice.

\begin{definition} Given a {lattice $(L, \wedge,\vee)$}, we say that an equivalence relation on $L$, which is compatible with both the supremum and the infimum of {$(L,\wedge, \vee)$} is a \emph{congruence} on $L$.
\end{definition}

Now, we introduce the notion of quotient lattice from a congruence based on the operations of the original lattice.

	\begin{definition}\label{def:latticecongruence}
		Given an equivalence relation $\theta$ on a lattice {$(L, \wedge,\vee)$}, two operators $\vee_\theta$ and $\wedge_\theta$ on the set of equivalence classes $L/\theta = \{[a]_{\theta} \mid a\in L\}$, for all $a, b\in L$, are defined as follows
		$$[a]_\theta \vee_\theta [b]_\theta = [a \vee b]_\theta \text{ and } [a]_\theta \wedge_\theta [b]_\theta = [a \wedge b]_\theta.$$
		$\vee_\theta$ and $\wedge_\theta$ are well defined on $L/\theta$ if and only if $\theta$ is a congruence.\\
		
		When $\theta$ is a congruence on $L$, we call $\left\langle L/\theta, \vee_\theta, \wedge_\theta\right\rangle $ \emph{the quotient lattice of $L$ modulo $\theta$}.
	\end{definition}

\newcommand{\pictcoveringB}{%
	\begin{picture}(1em,.5em)
	\roundcap
	\put(0,.25em){\line(1,0){.6em}}
	\put(.6em,.25em){\line(3,1){.4em}}
	\put(.6em,.25em){\line(3,-1){.4em}}
	\end{picture}%
}
\newcommand{\covering}{\mathrel{\text{$\vcenter{\hbox{\pictcoveringB}}$}}}

The following lemma {is} useful when calculating with congruences.
	
	\begin{lemma}[\cite{DaveyPriestley}]\label{lem:charact}
		Given a {lattice $(L, \wedge, \vee)$} we have that
		\begin{enumerate}[(i)]
			
			\item An equivalence relation $\theta$ on $L$ is a congruence if and only if, for all $a, b, c\in L$,
			$$(a, b) \in \theta \text{ implies }  (a\vee c, b \vee c) \in\theta  ~\text{ and }~ (a\wedge c, b \wedge c) \in\theta.$$
			\item Let $\theta$ be a congruence on $L$ and $a,b,c\in L$.
			\begin{enumerate}[(a)]
				\item If $(a, b) \in \theta$ and $a\preceq c \preceq b$, then $(a, c) \in \theta$.
				\item $(a, b) \in \theta$ if and only if $(a \wedge b, a \vee b) \in \theta$.
			\end{enumerate}
		
		\end{enumerate}
	\end{lemma}

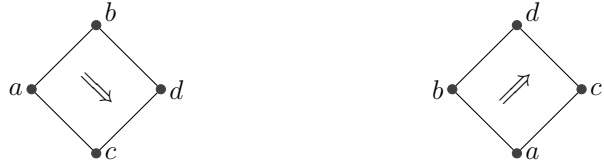
\begin{figure}[h!]
	
	\begin{minipage}{0.6\textwidth}
		\begin{center}
			\tikzstyle{place}=[circle,draw=black!75,fill=black!75]
			\tikzstyle{transition}=[circle,draw=black!75,fill=black!75]
			\begin{tikzpicture}
			[inner sep=0.5mm,scale=0.85, every node/.style={scale=0.85}]		
			\node at (0,0) (C0) [place, label=right:$c$] {};  			
			\node at (1,1) (C1) [transition, label=right:$d$] {};
			\node at (-1,1) (C2) [transition, label=left:$a$] {};
			\node at (0,2) (C3) [transition, label=5:$b$] {};  			
			
			\draw [-] (C0) -- (C1) -- (C3) ;
			\draw [-] (C0) -- (C2) -- (C3);
			\node at (0,1) [rotate=-45]{$\Longrightarrow$};					
			\end{tikzpicture}
		\end{center}
	\end{minipage}	
	\begin{minipage}{0.2\textwidth}
		\begin{center}
			\tikzstyle{place}=[circle,draw=black!75,fill=black!75]
			\tikzstyle{transition}=[circle,draw=black!75,fill=black!75]
			\begin{tikzpicture}
			[inner sep=0.5mm,scale=0.85, every node/.style={scale=0.85}]		
			\node at (0,0) (C0) [place, label=right:$a$] {};  			
			\node at (1,1) (C1) [transition, label=right:$c$] {};
			\node at (-1,1) (C2) [transition, label=left:$b$] {};
			\node at (0,2) (C3) [transition, label=5:$d$] {};  			
			
			\draw [-] (C0) -- (C1) -- (C3) ;
			\draw [-] (C0) -- (C2) -- (C3);
			\node at (0,1) [rotate=45]{$\Longrightarrow$};					
			\end{tikzpicture}
		\end{center}
	\end{minipage}

	\caption{Opposite sides of a quadrilateral.}
	\label{fig:lattice3}
\end{figure}

{The equivalence classes of a congruence are convex sublattices of the original lattice and besides are quadrilateral-closed. Let us recall the meaning of notion of quadrilateral-closed. Let $(L, \preceq)$ be a lattice, an equivalence relation $\theta$ on $(L, \preceq)$ and suppose that $\{a,b,c,d\}$ is a  subset of $L$ composed of four elements forming a quadrilateral, then $a,b$ and $c,d$ are said to be \emph{opposite sides of the quadrilateral} $\left\langle a,b; c,d\right\rangle$ (see Figure~\ref{fig:lattice3})  if $a\prec b, ~c\prec d$ and either:
			$$(a\vee d = b \text{ and } a\wedge d = c) ~\text{ or }~ (b\vee c = d \text{ and } b\wedge c = a).$$

Therefore, \emph{quadrilateral-closed} means that whenever given two opposite sides of a quadrilateral $a, b$ and $c, d$, satisfying that $a, b$ belong to an equivalence class, then $c, d$ belong to another or the same equivalence class, that is, if $a, b\in [x]_\theta$, with $x\in L$ then there exists $y\in L$ such that $c,d \in [y]_\theta$.}

The following result introduces an interesting characterization of congruence which will be fundamental for the purpose of this paper.
	
\begin{theorem}[\cite{DaveyPriestley}]\label{th:charact}
		Let {$(L,\wedge, \vee)$} be a lattice and let $\theta$ be an equivalence relation on $L$. Then $\theta$ is a congruence if and only if
		\begin{enumerate}[(i)]
			\item each equivalence class of $\theta$ is a sublattice of $L$,
			\item each equivalence class of $\theta$ is convex,
			\item the equivalence classes of $\theta$ are quadrilateral-closed.
		\end{enumerate}	
\end{theorem}

The set of congruences on a lattice $L$, denoted as Con $L$, is a topped $\cap$-structure on $L\times L$. Hence Con $L$, ordered by inclusion, is a complete lattice. The least element and the greatest element are given by $\theta_{\bot} = \{(a,a) \mid a\in L\}$ and $\theta_{\top} =\{(a,b) \mid a,b\in L\}$, respectively.\\

Given a lattice {$(L, \wedge, \vee)$} and two elements $a, b\in L$, the least congruence satisfying that $a$ and $b$ are related is denoted as $\theta_{(a,b)}$ and it is called the \emph{principal congruence generated by $(a,b)$} and it is defined as follows

		$$\theta_{(a,b)} = \bigwedge \{\theta \in\text{Con }L \mid (a,b)\in\theta\}.$$
	
Next lemma shows the importance of this definition.
	
\begin{lemma}[\cite{DaveyPriestley}]\label{Ch:princongruences}
		Let {$(L, \wedge, \vee)$} be a lattice and $\theta \in$ Con $L$. Then
		$$\theta = \bigvee \{\theta_{(a,b)} \mid (a,b)\in\theta\}.$$
\end{lemma}

Therefore, principal congruences factorize any congruence.

\subsection{Formal concept analysis}

Since equivalence relations will be considered in this work to reduce concept lattices, basic definitions of FCA are recalled in order to understand the motivation and results presented in this paper. 

\begin{definition}
	A \emph{context} is a triple $(A,B,R)$ with a set of attributes $A$, a set of objects $B$ and a crisp relationship $R\subseteq A\times B$. We will write $R(a,b) = 1$ when $(a,b)\in R$ and $R(a,b)=0$ when $(a,b)\notin R$.
\end{definition}

Furthermore, if we consider a context, two mappings, ${\ }^\uparrow\colon 2^B\to 2^A$ and ${\ }^\downarrow\colon 2^A\to 2^B$, can be  defined for each $X\subseteq B$ and $Y\subseteq A$  as:
\begin{eqnarray}\label{def-classical.gc1}
X^{\uparrow}&=&\{a\in A\mid (a,x)\in R, \hbox{for all } x\in X\}\label{def-classical.gc2}\\
Y^{\downarrow}&=&\{x\in B\mid (a,x)\in R, \hbox{for all } a\in Y\}
\end{eqnarray}
These operators form a Galois connection~\cite{DaveyPriestley}, which leads us to the following definition.

\begin{definition}
Given a context $(A,B,R)$ and the operators ${\ }^\uparrow$ and ${\ }^\downarrow$ defined above. If for a pair $(X,Y)$ with $X \subseteq B$ and $Y \subseteq A$, the equalities $X^{\uparrow}=Y$ and $Y^\downarrow=X$ hold, then the pair $(X,Y)$ is called \emph{concept}. 
\end{definition} 

Given a pair of concepts $(X_1,Y_1)$ and $(X_2,Y_2)$, we say that $(X_1,Y_1) \leq (X_2,Y_2)$ if $X_1 \subseteq X_2$ ($Y_2\subseteq Y_1$, equivalently). The set of all concepts with this ordering relation has the structure of a complete lattice, it is called \emph{formal concept lattice} and it is denoted as $\mathcal C(A,B,R)$~\cite{DaveyPriestley,GanterW}.

Now, we recall two results about reduction in FCA~\cite{bmrd:RSTFCA:c}.
The first one shows that when we reduce the set of attribute of a formal context, an equivalence relation on the set of concepts of the original concept lattice is induced.		

\begin{proposition}[\cite{bmrd:RSTFCA:c}]\label{prop:clase2}
	Given a context $(A, B, R)$ and a subset $D \subseteq {A}$.	
	The set $ {\rho_D}=\{((X_1,Y_1),(X_2,Y_2)) \mid (X_1,Y_1),(X_2,Y_2)\in \mathcal C(A,B, R), X_1^{\uparrow_D\downarrow}= X_2^{\uparrow_D\downarrow}\}$ is an equivalence relation. Where ${}^{\uparrow_D}$ denotes the concept-forming operator, given in Expresion~\eqref{def-classical.gc2}, restricted to the subset of attributes	${D\subseteq A}$.
\end{proposition}

The next result shows that every class of the equivalence relation defined above has the structure of a join semilattice with maximum element.

\begin{proposition}[{\cite{bmrd:RSTFCA:c}}]\label{prop:clase3}
	Given a context $(A, B, R)$, a subset $D \subseteq {A}$ and a class $[(X,Y)]_D$ of the quotient set $ \mathcal C(A,B, R)/{\rho_D}$. The class $[(X,Y)]_D$  is a join semilattice with maximum element ${(X^{\uparrow_D\downarrow},X^{\uparrow_D\downarrow\uparrow}})$.
\end{proposition}

Hence, we cannot ensure that the classes are sublattices of the original concept lattice, as it was shown in  Example 3.10 of~\cite{bmrd:RSTFCA:c}. Therefore, it is interesting to study when these classes are sublattices and the properties of the obtained reduction. {These results have been extended to the fuzzy FCA framework of multi-adjoint concept lattices in~\cite{bmrd:FCARST:f}. This framework  was introduced by Medina, Ojeda-Aciego  and Ruiz-Calvi{\~{n}}o in~\cite{mor-fss-cmpi} with the main goal of presenting a general and flexible FCA framework based on the multi-adjoint philosophy.  Multi-adjoint concept lattice generalizes different fuzzy extension of FCA~\cite{ludomir14,BelohlavekB05,burusco:1998} and has  widely been  studied   in diverse papers~\cite{TFS:2020-acmr,Cornejo2017,ins2018:cmr}. See the basic notions in~\cite{mor-fss-cmpi}. 

In the multi-adjoint concept lattice framework a multi-adjoint frame $(L_1,L_2,P, \adjoint_1,
	\dots,\adjoint_n  )$  needs to be fixed on which a   context $(A,B,R,\sigma)$  is defined and a concept lattice $\mathcal{M}(A, B, R, \sigma)$ is obtained.  On this framework,  the authors   in~\cite{bmrd:FCARST:f} also proved}
that  a reduction of the set of attributes induces an equivalence relation on $\mathcal{M}(A, B, R, \sigma)$, in which the equivalence classes are join-subsemilattices. This result will be recalled next, where  $\uparrow_D$ and  $\downarrow^D$ are the concept-forming operators  associated with the subcontext $ \mathcal M(D,B, R_{|D\times B},\sigma)$, with $D\subseteq A$.

\begin{proposition}[\cite{bmrd:FCARST:f}]\label{prop:partition}	
	Let $D \subseteq {A}$ be a subset of attributes.
	The set $ {\rho_D} =\{(\langle g_1,f_1\rangle,\langle g_2,f_2\rangle) \mid \langle g_1,f_1\rangle,\langle g_2,f_2\rangle\in \mathcal M(A,B, R,\sigma), g_1^{\uparrow_D\downarrow^D}= g_2^{\uparrow_D\downarrow^D}\}$ is an equivalence relation and every class $[\langle g,f\rangle]_D$ of $ \mathcal M(A,B, R,\sigma)/{\rho_D}$   is a join-semilattice with maximum element $\langle g^{\uparrow_D\downarrow^D},g^{\uparrow_D\downarrow^D\uparrow}\rangle$.
\end{proposition}

Once we have recalled  the  required  preliminary  notions, the main contributions of this work are presented in the following section.

\section{Weakening the notion of congruence}\label{def:wc:red}

In this section, we present a weaker notion of congruence relation in order to complement reduction mechanisms in FCA. As we have recalled,  any reduction of the set of attributes generates a partition in the set of concepts associated with the original context, where the obtained equivalence classes may not form sublattices of the original concept lattice. Our interest lies in generating groups of concepts with a closed algebraic structure by complementing the  reductions given in FCA. 

This goal can be achieved through the notion of congruence relation on lattices. Therefore, we will consider the use of congruence relations to reduce  concept lattices, and we will analyze the obtained results. In particular, we are interested in the least congruence whose equivalence classes contain the equivalence classes induced by an attribute reduction of a context. Usually, this reduction is given by reducts which are minimal subsets of attributes preserving the information in the dataset. More details are included in~\cite{bmrd:RSTFCA:c}.

Next, we illustrate the result of applying congruences through a practical example considered in~\cite{bmrd:RSTFCA:c}. {In this example, we will show that the equivalence classes induced by a reduction procedure  may be noticeably different from the ones provided by the least congruence containing the  partition induced by  the reduction, as a consequence, this difference would entail a relevant loss of information}.

\begin{example}\label{Ex1}
	Given the formal context $({A}, B, R)$ displayed in Table~\ref{tabla:planets}, where the set of objects in $B$ are the planets of the Solar System together with the dwarf planet Pluto, that is $B=\{$Mercury~(M), Venus~(V), Earth~(E), Mars~(Ma), Jupiter~(J), Saturn~(S), Uranus~(U), Neptune~(N), Pluto~(P)$\}$ and the set of attributes $A = \{$\tss\, (\mss), \tms\, (\ms), \tls\, (\ls), \tns\, (\ns), \tfs, (\fs), \tmy\, (\my), \tmn\, (\mn)$\}$. 
	
	\setlength{\tabcolsep}{0.25cm}				
	\begin{table}[h]
		\begin{center}
			\begin{tabular}{l|c c c c c c c c c}
				\hline
				$R$ & M  & V & E & Ma & J & S & U & N & P\\ \hline
				$\tss$ & 1 & 1 & 1 & 1 & 0 & 0 & 0 & 0 & 1\\ 
				$\tms$ & 0 & 0 & 0 & 0 & 0 & 0 & 1 & 1 & 0\\
				$\tls$ & 0 & 0 & 0 & 0 & 1 & 1 & 0 & 0 & 0\\
				$\tns$ & 1 & 1 & 1 & 1 & 0 & 0 & 0 & 0 & 0\\
				$\tfs$ & 0 & 0 & 0 & 0 & 1 & 1 & 1 & 1 & 1\\
				$\tmy$ & 0 & 0 & 1 & 1 & 1 & 1 & 1 & 1 & 1\\
				$\tmn$ & 1 & 1 & 0 & 0 & 0 & 0 & 0 & 0 & 0\\\hline
			\end{tabular}
		\end{center}
		\caption{Relation of Example~\ref{Ex1}.}
		\label{tabla:planets}
	\end{table}	
	
	In the left side of \refig{Ex1:latticeRS}, it is displayed the concept lattice from the given context. In~\cite{bmrd:RSTFCA:c},  the rough set reduct $D_1=\{\tss,\tms,\\ \tns,\tmy\}$ was considered to reduce the context. According to Proposition~\ref{prop:clase2}, this reduction makes that the concepts of the original concept lattice are grouped in equivalence classes which are represented in the middle of \refig{Ex1:latticeRS} by means of a Venn diagram.
	We know that each equivalence class has the structure of a join semilattice with maximum element as it is stated in Proposition~\ref{prop:clase3}. 
	\begin{figure}[h!]
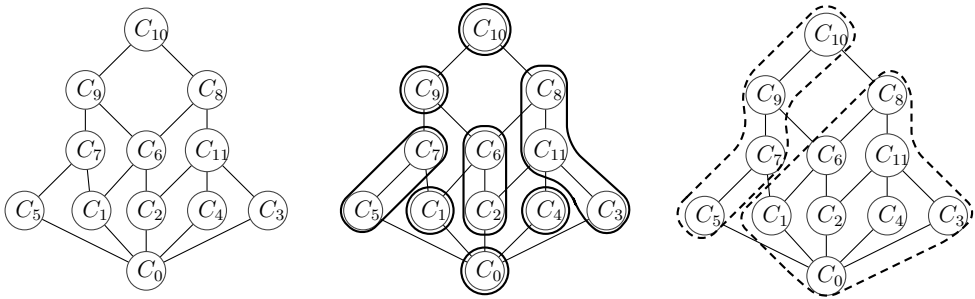

		\minipagethree{
			\figlat{}
		}{
			\figlat{\draw[thick] (C1) circle (11pt);
				\draw[thick] (1,1) circle (11pt);
				\draw[thick] (C9) circle (11pt);
				\draw[thick] (C10) circle (12pt);
				\draw[thick] (C0) circle (11pt);
				\draw[ rounded corners=0.25cm,thick] (-0.35,0.6) rectangle (0.35,2.4);
				\draw[rotate around={45:(-2,0.5)}, rounded corners=0.25cm,thick] (-2,0.52) rectangle (0.15,1.2);
				\draw[-,rounded corners=0.25cm,thick,cap=round, join=round] (1.4,1.4)-- ++(-0.8,0.3)-- ++(0,1.7)-- ++(0.75,0)-- ++(0,-1.3)-- ++(1.15,-1.05)-- ++(-0.45,-0.55)-- ++ (-0.5,0.4)-- cycle;}
		}{
			\figlat{\draw[-,densely dashed,rounded corners=0.25cm,thick] (-2.3,1.3)-- ++(1,1)-- ++(-0.1,0.9)-- ++(1.4,1.4)-- ++(0.6,-0.6)-- ++(-1.3,-1.2)-- ++(0.1,-0.85)-- ++(-1.45,-1.45)-- ++(-0.525,0.525)-- ++(0.3,0.3);
				\draw[-,densely dashed,rounded corners=0.2cm,thick] (-0.5,-0.05)-- ++(-1,1)-- ++(2.55,2.55)-- ++(0.35,-0.35)
				-- ++(0,-1)-- ++(1.1,-1.1)-- ++(-0.4,-0.4)-- ++(-2.15,-1.05)-- cycle;}
		}				
		\caption{The original concept lattice (left),  the  obtained reduction in~\cite{bmrd:RSTFCA:c} (middle) and the least congruence containing the previously reduction (right).}
		\label{Ex1:latticeRS}
	\end{figure}
	
	We can bring together the reduction given in FCA and congruences, 
	finding the least congruence such that each equivalence class induced by the reduction of the context is  included  in one equivalence class provided by the congruence relation. This least congruence is shown in the right side of \refig{Ex1:latticeRS} by means of a dashed Venn diagram. As we can see in \refig{Ex1:latticeRS}, this congruence relation is composed of only two equivalence classes, since it has grouped too many concepts in each class. Consequently, in this case, the use of congruences entails a relevant loss of information, which is not convenient in any process of data analysis. \qed
\end{example}
	
	The result obtained in the previous example reveals the necessity of a  weaker notion of congruence removing the quadrilateral-closed property and preserving the other two properties in the characterization given in Theorem~\ref{th:charact}. This weaker notion is introduced  in the following definition. 
		
	\begin{definition}\label{def:lc}
		Given a lattice $\left( L, \preceq \right)$, we say that an equivalence relation $\delta$ on $L$ is a \emph{local congruence} if the following properties hold:
		
		\begin{enumerate}
			\item[(i)] each equivalence class of $\delta$ is a sublattice of $L$,
			\item[(ii)] each equivalence class of $\delta$ is convex.
		\end{enumerate}	
	\end{definition}

{
\begin{remark}\label{remark:ex1}
Clearly, the attribute reduction of the concept lattice introduced in Example~\ref{Ex1} provides a local congruence (concept lattice in the middle of Figure~\ref{Ex1:latticeRS}). Therefore, the introduced notion offers a better reduction than the one provided using congruences, aggregating as less concepts (information) as possible. Moreover, since in this particular case the reduction already produces equivalence classes that are convex sublattices, then   the amount of lost information with this weaker notion of congruence is minimized as much as possible.
\end{remark}}

Although this new definition is a weak definition of the notion of congruence, the name of weak-congruence has already been used in the literature~\cite{Vojvodic1988,Tepavcevic2015,Tepavcevic2008,Walendziak2002}
in order to define congruences without the reflexivity property, that is, a weak congruence is a symmetry, transitivity and compatible relation. Therefore,  another suitable name has been considered in this paper to the introduced  general notion of congruence, whose justification will be introduced after the next direct characterization of Definition~\ref{def:lc}
 in terms of the  equivalence relation $\delta$.

\begin{proposition}\label{prop:ch-relation}
 Given a lattice $( L, \preceq)$ and an equivalence relation $\delta$ on $L$, the relation $\delta$ is a local congruence on $L$ if and only if, for each $a, b, c \in L$, the following properties hold:
 \begin{enumerate}
 	\item[(i)] If $(a,b)\in\delta$ and $a\preceq c\preceq b$, then $(a,c)\in\delta$.
 	\item[(ii)] $(a,b)\in\delta$ if and only if $(a\wedge b, a\vee b)\in\delta$.
 \end{enumerate}
 \end{proposition}
\begin{proof}
The proof holds directly from the definition of local congruence.\qed
\end{proof}

This result will be basic to introduce  the following characterization, which generalizes Lemma~\ref{lem:charact}(i) and motivates the considered notion. 
\begin{proposition}\label{prop:charact}
Given a lattice $(L, \preceq)$ we have that an equivalence relation $\delta$ on $L$ is a local congruence if and only if, for all $a, b,c\in L$, if $(a, b) \in \delta$ and $a\wedge b \preceq c \preceq a\vee b$,  then we have that 
$$  (a\vee c, b \vee c) \in\delta  ~\text{ and }~ (a\wedge c, b \wedge c) \in\delta$$
\end{proposition}

\begin{proof}
Let us assume that $\delta$ is a local congruence on $L$ and we  consider $a, b, c\in L$ such that $(a,b)\in \delta$ and $a\wedge b \preceq c \preceq a\vee b$. Straightforwardly, by Proposition~\ref{prop:ch-relation}, we have that
\begin{equation}\label{P2}
(a,a\vee b)\in\delta ~\text{and}~(b,a\vee b)\in\delta.
\end{equation}

In addition, since $a\wedge b \preceq c \preceq a\vee b$, by the supremum property,  the following inequalities hold 
$$a\vee (a\wedge b) \preceq a\vee c \preceq a \vee (a\vee b),$$
which is equivalent to
$$ a \preceq a\vee c \preceq a\vee b.$$
Hence, as $(a,a\vee b)\in\delta$, by Proposition~\ref{prop:ch-relation}(i), we obtain that $(a, a\vee c)\in \delta$. 
Considering an analogous procedure to the previous one, we have that $(b, b\vee c)\in \delta$.
Therefore, since $(a, a\vee c)\in \delta$ and $(b, b\vee c)\in \delta$, by the hypothesis $(a,b)\in \delta$ and the transitivity property of $\delta$, we can assert that $(a\vee c, b \vee c) \in\delta$. Analogously, we have that $(a\wedge c, b \wedge c) \in\delta$.

Now, let us assume that $\delta$ is an equivalence relation on $L$, such that, if  $(a,b)\in \delta$ and $a\wedge b \preceq c \preceq a\vee b$, then it satisfies that 
$$ 
(a\vee c, b \vee c) \in\delta ~\text{and}~ (a\wedge c, b \wedge c) \in\delta
$$
for all $a, b, c\in L$.
   
We will use Proposition~\ref{prop:ch-relation} in order to prove that $\delta$ is a local congruence. 
If   $(a,b)\in \delta$ and    $a\preceq c\preceq b$ then, in particular,  $a\wedge b \preceq a\preceq c \preceq b\preceq a\vee b$. Therefore, by hypothesis, we have that $(a\wedge c, b \wedge c)\in\delta$. Since $a\wedge c = a$ and $c\wedge b = c$,  we obtain that $(a,c)\in\delta$.
 Hence,  item $(i)$ of Proposition~\ref{prop:ch-relation} is satisfied.

In addition, for each $(a,b)\in \delta$ if we consider $c = a$, by hypothesis,  we have that $(a\vee c, b\vee c)\in\delta$, that is, $(a, b\vee a)\in\delta$. Similarly, for $c = b$ we have that $(a\wedge b, b)\in\delta$. Hence, by the transitivity property of $\delta$, we obtain that $(a\wedge b, a\vee b)\in\delta$. Following a similar reasoning, we can easily prove that, for all $a,b\in L$, if $(a\wedge b, a\vee b)\in\delta$ then $(a,b)\in \delta$. Therefore, item $(ii)$ of Proposition~\ref{prop:ch-relation} also holds. 

Consequently, we can conclude that $\delta$ is a local congruence on $L$.\qed
\end{proof}

Hence, the difference from the equivalence given in Lemma~\ref{lem:charact}(i) is that in the aforementioned lemma, the element $c$ is arbitrary in $L$, and in Proposition~\ref{prop:charact} is a \emph{local} element bounded by $a\wedge b$ and $a\vee b$.

As we highlighted above, the particular algebraic structure of the equivalence classes is the most important property associated with the new notion. The set of these classes is formally defined next.

	\begin{definition}
			Let  $( L, \preceq)$ be a lattice and $\delta$  a local congruence, the  quotient set $L/\delta$ provides a partition of $L$, which is called \emph{local congruence partition} (or \textit{lc-partition} in short) of $L$ and it is denoted as~$\pi_\delta$.  The elements in the lc-partition $\pi_\delta$ are convex sublattices of $L$.
	\end{definition}

	According to the previous definition, it can be noted that each equivalence class of the quotient set $L/\delta$ is a closed algebraic structure. Moreover, each local congruence relation univocally determines a local congruence partition and vice versa. As a consequence, both notions can be considered indistinctly.

In the following section, a formal definition of ordering among the classes of the quotient set of a local congruence will be studied.

\section{The quotient set of a local congruence}\label{sec:poset}
Now, we focus on the equivalence classes of a quotient set provided by a local congruence. We are interested in studying how we can establish an ordering relation between these classes. The following definition will play a key role for this purpose.

\begin{definition}\label{Def:sequence_cycle}
	Let $(L,\preceq)$ be a lattice and a local congruence $\delta$ on $L$.
	\begin{enumerate}[(i)]
		\item A sequence of elements of $L$, $(p_0, p_1, \dots, p_{n})$ with ${n\geq 1}$, is called a \emph{$\delta$-sequence}, denoted as $(p_0, p_n)_\delta$, if for each $i \in \{1,\dots,n\}$ either $(p_{i-1},p_i)\in \delta$ or $p_{i-1}\preceq p_i $ holds.
		\item If a $\delta$-sequence, $(p_0, p_n)_\delta$, satisfies that $p_0 = p_n$, then it is called a \emph{$\delta$-cycle}. In addition, if the $\delta$-cycle satisfies that $[p_0]_\delta = [p_1]_\delta = \dots = [p_n]_\delta$ is said to be \emph{closed}.
	\end{enumerate} 
\end{definition}

The notions in Definition \ref{Def:sequence_cycle} are clarified in \refig{Ex:dsequence} and \refig{Ex:dcycle}, where the triple vertical line that connects $p_{i-1}$ with $p_i$ means that they are related under the considered local congruence relation, that is, $(p_{i-1},p_i)\in \delta$. The simple line indicates that the two elements are connected by means of the ordering relation defined on the lattice.

The following definition provides a first step to define a partial order on the quotient set provided by a local congruence.

\begin{definition}\label{Def:preorden}
	Given a lattice $(L,\preceq)$ and a local congruence $\delta$ on $L$, we define a binary relation $\preceq_\delta$ on $L/\delta$ as follows:
	\begin{center}
		$[x]_\delta \preceq_\delta [y]_\delta$\quad if there exists a $\delta$-sequence $(x', y')_\delta$
	\end{center}
	for some $x'\in [x]_\delta$ and $y'\in[y]_\delta$.
\end{definition}
	\begin{figure}[h!]
	\begin{center}
		\tikzstyle{place}=[circle,draw=black!75,fill=black!75]
		\tikzstyle{transition}=[circle,draw=black!75,fill=black!75]
		\tikzstyle{point} = [circle,draw=white,fill=white]
		\begin{tikzpicture}[inner sep=0.75mm,scale=0.9, every node/.style={scale=0.9}]		
		\tikzset{
			triple/.style args={[#1] in [#2] in [#3]}{
				#1,preaction={preaction={draw,#3},draw,#2}
			}
		} 				
		\node at (-5,0) (a1) [place, label=270:$p_0$] {};  			
		\node at (-3,0) (a2) [place, label=270:$p_2$] {};
		\node at (-3,1.5) (b2) [transition, label=above:$p_1$] {};
		\node at (-1,0) (a3) [place, label=270:$p_4$] {};
		\node at (-1,1.5) (b3) [transition, label=above:$p_3$] {};
		\node at (1.7,0) (an) [place, label=270:$p_{n-1}$] {};
		\node at (1.7,1.5) (bn) [transition, label=above:$p_{n-2}$] {};
		\node at (3.7,1.5) (bn1) [place, label=above:$p_{n}$] {};
		\node at (0.5,0.75) (p) [point] {$\dots$};
		
		\draw [-] (a1) -- (b2);
		\draw [-] (a2) -- (b3);
		\draw [-] (a3) -- (0,0.75);
		\draw [-] (0.8,0.75) --(bn);
		\draw [-] (an) -- (bn1);
		
		\draw [triple={[line width=0.3mm,black] in
			[line width=0.9mm,white] in
			[line width=1.5mm,black]}] (-3,0.15) to (-3,1.35);
		\draw [triple={[line width=0.3mm,black] in
			[line width=0.9mm,white] in
			[line width=1.5mm,black]}] (-1,0.15) to (-1,1.35);
		\draw [triple={[line width=0.3mm,black] in
			[line width=0.9mm,white] in
			[line width=1.5mm,black]}] (-3,0.15) to (-3,1.35);
		\draw [triple={[line width=0.3mm,black] in
			[line width=0.9mm,white] in
			[line width=1.5mm,black]}] (1.7,0.15) to (1.7,1.35);
		\end{tikzpicture}
	\end{center}
	\caption{Example of $\delta$-sequence.}
	\label{Ex:dsequence}
\end{figure}
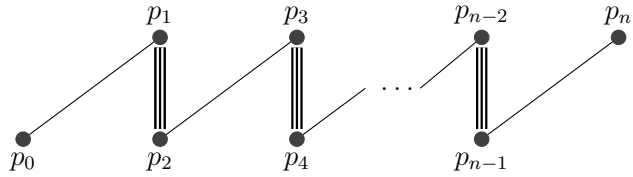

\begin{figure}[h!]
	\begin{center}
		\tikzstyle{place}=[circle,draw=black!75,fill=black!75]
		\tikzstyle{transition}=[circle,draw=black!75,fill=black!75]
		\tikzstyle{point} = [circle,draw=white,fill=white]
		\begin{tikzpicture}[inner sep=0.75mm,scale=0.9, every node/.style={scale=0.9}]		
		\tikzset{
			triple/.style args={[#1] in [#2] in [#3]}{
				#1,preaction={preaction={draw,#3},draw,#2}
			}
		} 				
		\node at (-5,1.5) (b1) [place, label=above:${p_0 = p_n}$] {};
		\node at (-5,0) (a1) [place, label=270:$p_1$] {};  			
		\node at (-3,0) (a2) [place, label=270:$p_3$] {};
		\node at (-3,1.5) (b2) [transition, label=above:$p_2$] {};
		\node at (-1,0) (a3) [place, label=270:$p_5$] {};
		\node at (-1,1.5) (b3) [transition, label=above:$p_4$] {};
		\node at (2,0) (an) [place, label=270:$p_{n-1}$] {};
		\node at (2,1.5) (bn) [transition, label=above:$p_{n-2}$] {};
		\node at (0.5,0.75) (p) [point] {$\dots$};
		
		\draw [-] (a1) -- (b2);
		\draw [-] (a2) -- (b3);
		\draw [-] (a3) -- (0,0.75);
		\draw [-] (0.9,0.75) --(bn);
		\draw [-] (an) -- (b1);
		
		\draw [triple={[line width=0.3mm,black] in
			[line width=0.9mm,white] in
			[line width=1.5mm,black]}] (-3,0.15) to (-3,1.35);
		\draw [triple={[line width=0.3mm,black] in
			[line width=0.9mm,white] in
			[line width=1.5mm,black]}] (-1,0.15) to (-1,1.35);
		\draw [triple={[line width=0.3mm,black] in
			[line width=0.9mm,white] in
			[line width=1.5mm,black]}] (-3,0.15) to (-3,1.35);
		\draw [triple={[line width=0.3mm,black] in
			[line width=0.9mm,white] in
			[line width=1.5mm,black]}] (2,0.15) to (2,1.35);
		\draw [triple={[line width=0.3mm,black] in
			[line width=0.9mm,white] in
			[line width=1.5mm,black]}] (-5,0.15) to (-5,1.35);
		\end{tikzpicture}
	\end{center}
	\caption{Example of  $\delta$-cycle.}
	\label{Ex:dcycle}
\end{figure}
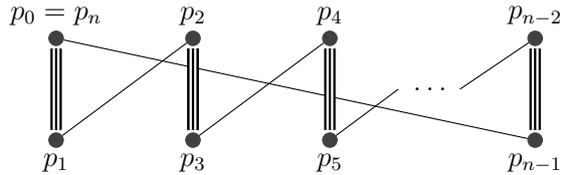

Notice that the relation $\preceq_\delta$ given in Definition~\ref{Def:preorden} is a preorder. Clearly, by  definition, $\preceq_\delta$ is reflexive and transitive.
However, the relation $\preceq_\delta$ is not a partial order since the antisymmetry property does not hold, for any local congruence in general.  In the following example, we show a case in which the previously defined relation  $\preceq_\delta$ does not satisfy the antisymmetry property.

\begin{example}
\label{orden_no_antisim}	
Given the lattice $(L,\preceq)$ and  the local congruence $\delta$ given in \refig{Contraejemplo}. 
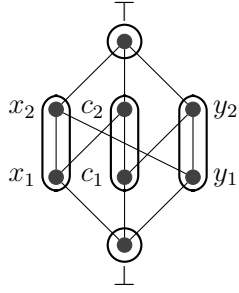
\begin{figure}[h!]		
	\begin{center}
		\tikzstyle{place}=[circle,draw=black!75,fill=black!75]
		\begin{tikzpicture}[inner sep=0.75mm,scale=0.9, every node/.style={scale=0.9}]		
		\node at (0,0) (0) [place, label={[label distance=0.1cm]below:$\bot$}] {};
		\node at (-1,1) (a) [place, label={[label distance=0.1cm]left:$x_1$}] {};  			
		\node at (0,1) (b) [place, label={[label distance=0.1cm]left:$c_1$}] {};
		\node at (1,1) (c) [place, label={[label distance=0.1cm]right:$y_1$}] {};
		\node at (-1,2) (a1) [place, label={[label distance=0.1cm]left:$x_2$}] {};  			
		\node at (0,2) (b1) [place, label={[label distance=0.1cm]left:$c_2$}] {};
		\node at (1,2) (c1) [place, label={[label distance=0.1cm]right:$y_2$}] {};
		\node at (0,3) (1) [place, label={[label distance=0.1cm]above:$\top$}] {};
		
		\draw [-] (0) -- (a)-- (a1)--(1);
		\draw [-] (0) -- (b)-- (b1)--(1);
		\draw [-] (0) -- (c)-- (c1)--(1);
		\draw [-] (a) -- (b1);
		\draw [-] (b) -- (c1);
		\draw [-] (c) --(a1);
		
		\draw[ rounded corners=0.2cm,thick] (-1.2,0.8) rectangle (-0.8,2.2);
		\draw[ rounded corners=0.2cm,thick] (-0.2,0.8) rectangle (0.2,2.2);
		\draw[ rounded corners=0.2cm,thick] (0.8,0.8) rectangle (1.2,2.2);
		\draw[thick] (0,0) circle (7pt);
		\draw[thick] (0,3) circle (7pt);
		\end{tikzpicture}
	\end{center}
	\caption{Example where $\preceq_\delta$ is not a partial order.}
	\label{Contraejemplo}	
\end{figure}
We have
that the equivalence classes of $\delta$ are $[\top]_\delta = \{\top\}, [x_1]_\delta =\{x_1, x_2\}, [c_1]_\delta= \{c_1, c_2\}, [y_1]_\delta = \{y_1, y_2\}$ and $[\bot]_\delta = \{\bot\}$. It is easy to check that these equivalence classes are convex sublattices of $L$. Moreover, we can observe that $[x_1]_\delta \preceq_\delta [y_1]_\delta$ since there exists a $\delta$-sequence, $(x_1, y_2)_\delta =( x_1, c_2, c_1, y_2)$, and also $[y_1]_\delta \preceq_\delta [x_1]_\delta$ since there also exists a $\delta$-sequence, $(y_1,x_2)_\delta=( y_1, x_2)$, but $[x_1]_\delta \neq [y_1]_\delta$ and thus $\preceq_\delta$ is not antisymmetric.\qed
\end{example}

There are certain cases in which the relation $\preceq_\delta$ is a partial order, depending on the local congruence.

\begin{example}
	\label{orden_antisim}
	Let $(L,\preceq)$  be a lattice isomorphic to the concept lattice given in Example~\ref{Ex1} and $\delta$  the local congruence shown in the left side of \refig{ex:wc-reduction}. In this case, the considered local congruence makes that the relation $\preceq_\delta$  satisfies the antisymmetry property and, consequently, the relation $\preceq_\delta$ is a partial order and the Hasse diagram of $(L/\delta, \preceq_\delta)$ can be given (right side of Figure~\ref{ex:wc-reduction}).\qed
	
	\begin{figure}[h!]
	\begin{center}
		\minipagetwo{
			\figlatnodes{\draw[rotate around={135:(-1.5,1.5)},thick] (-1.4,1.5) ellipse (10pt and 35pt);
				\draw[thick] (0,1.5) ellipse (12pt and 28pt);
				\draw[thick] (-1.1,1) circle (9pt);
				\draw[thick] (1.1,1) circle (10pt);
				\draw[thick] (-1,3) circle (10pt);
				\draw[thick] (0,4.1) circle (11pt);
				\draw[thick] (0,-0.1) circle (11pt);
				
				\draw[-,rounded corners=0.2cm,thick] (2.1,0.6)-- ++(-1.35,1.35)-- ++(0,1.3);
				\draw[-,rounded corners=0.2cm,thick] (1.4,3.2)-- ++(0,-1.1)-- ++(1.1,-1.1);
				\draw[thick] (1.4,3.2) arc (0:180:3.25mm);
				\draw[thick] (2.1,0.6) arc (225:360:2.85mm) arc (0:50:2.85mm);
				\draw[->] (3,2) -- node[above]{$\preceq_\delta$} (6,2);
		}}
		{\tikzstyle{place}=[circle,draw=black!75,fill=black!75]
			\tikzstyle{transition}=[circle,draw=black!75,fill=black!75]
			\begin{tikzpicture}[inner sep=0.75mm,scale=0.75, every node/.style={scale=0.75}]	
			\node at (0,0) (C0) [place, label=below:$\bot$] {};  			
			\node at (0,2) (C2) [transition, label=below:$g'$] {};
			\node at (1,1) (C4) [transition, label=right:$d$] {};
			\node at (-1,1) (C1) [transition, label=left:$b$] {};
			\node at (-2,2) (C5) [transition, label=right:$f'$] {};
			\node at (1,3) (C8) [transition, label=right:$j'$] {};
			\node at (-1,3) (C9) [transition, label=left:$i$] {};
			\node at (0,4) (C10) [transition, label=above:$\top$] {};
			
			\draw [-] (C1) -- (C0);
			\draw [-] (C4) -- (C0);
			\draw [-] (C8) -- (C4);
			\draw [-] (C2) -- (C1);
			\draw [-] (C8) -- (C2);
			\draw [-] (C5) -- (C1);
			\draw [-] (C9) -- (C2);
			\draw [-] (C9) -- (C5);
			\draw [-] (C10) -- (C8);
			\draw [-] (C10) -- (C9);
			
			\end{tikzpicture}
		}
		\caption{A local congruence $\delta$ on $L$ (left) and its quotient set $(L/\delta, \preceq_\delta)$ (right).}
		\label{ex:wc-reduction}
	\end{center}
\end{figure}
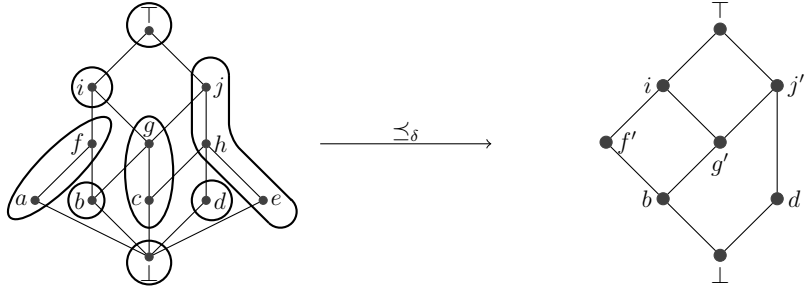

\end{example}

The following  results state  different conditions under which the relation $\preceq_\delta$ is a partial order.
\begin{proposition}\label{P:partialorder}
	
		Given a lattice $(L,\preceq)$ and a local congruence $\delta$, if for any two equivalence classes $[x]_\delta, [y]_\delta \in L/\delta$ there exists only one class $[c]_\delta\in L/\delta$ such that $[x]_\delta \preceq_\delta [c]_\delta \preceq_\delta [y]_\delta$ and $[y]_\delta \preceq_\delta [c]_\delta \preceq_\delta [x]_\delta$ satisfying that $x_1 \preceq c_1 \preceq y_1$ and $y_2 \preceq c_2 \preceq x_2$ with $x_1,x_2\in [x]_\delta$, $c_1,c_2\in [c]_\delta$ and $y_1, y_2\in [y]_\delta$, then $[x]_\delta = [y]_\delta$.
\end{proposition}

\begin{proof}

	Let us consider  two equivalence classes $[x]_\delta, [y]_\delta \in L/\delta$. Hence, there exists a class $[c]_\delta\in L/\delta$ such that $[x]_\delta \preceq_\delta [c]_\delta \preceq_\delta [y]_\delta$ and $[y]_\delta \preceq_\delta [c]_\delta \preceq_\delta [x]_\delta$ satisfying that $x_1 \preceq c_1 \preceq y_1$ and $y_2 \preceq c_2 \preceq x_2$ with $x_1,x_2\in [x]_\delta$, $c_1,c_2\in [c]_\delta$ and $y_1, y_2\in [y]_\delta$. Then, we have that $[x]_\delta \preceq_\delta [c]_\delta$ and $[c]_\delta \preceq_\delta [x]_\delta$ and, since the classes of $\delta$ are sublattices of $L$, we also have that $x_1 \vee x_2$ and $c_1 \vee c_2$ exist and belong to the classes $[x]_\delta$ and $[c]_\delta$, respectively. In addition, we have that $x_1 \vee c_2 \preceq c_1 \vee c_2$ and $x_1 \vee c_2 \preceq x_1 \vee x_2$, thus $x_1 \vee c_2 \preceq (x_1 \vee x_2) \wedge (c_1 \vee c_2)$. Hence, $x_1 \preceq (x_1 \vee x_2) \wedge (c_1 \vee c_2) \preceq x_1 \vee x_2$ and $c_2 \preceq (x_1 \vee x_2) \wedge (c_1 \vee c_2) \preceq c_1 \vee c_2$, by the convexity of the classes we have that $(x_1 \vee x_2) \wedge (c_1 \vee c_2)$ belongs to both classes, which implies that both classes are just the same class: $[x]_\delta = [c]_\delta$.
	
	We can proceed in an analogous way in order to prove that  $[c]_\delta=[y]_\delta$. Hence, we obtain that $[x]_\delta = [c]_\delta = [y]_\delta$. \qed
\end{proof}

From the previous proposition we obtain the following corollary.

\begin{corollary}
	Given a lattice $(L,\preceq)$ and a local congruence $\delta$, if for any two equivalence classes $[x]_\delta, [y]_\delta \in L/\delta$ such that $[x]_\delta \preceq_\delta [y]_\delta$ and $[y]_\delta \preceq_\delta [x]_\delta$ satisfy that $x_1 \preceq y_1$  and $y_2 \preceq x_2$ with $x_1,x_2\in [x]_\delta$ and $y_1, y_2\in [y]_\delta$, then $[x]_\delta = [y]_\delta$.
\end{corollary}

\begin{proof}
	
	It is straightforwardly deduced  from Proposition~\ref{P:partialorder} taking the class $[c]_\delta$ as either the class $[x]_\delta$ or $[y]_\delta$.	
 \qed
\end{proof}

It is easy to see in Figure~\ref{Contraejemplo} that $\preceq_\delta$ is not a partial order because of there exists a $\delta$-cycle composed of elements belonging to different equivalence classes, i.e. the $\delta$-cycle $(x_2,x_1,c_2,c_1,y_2,y_1,x_2)$. In order to avoid this problem, every  $\delta$-cycle must be contained in one single class, that is, every $\delta$-cycle must be closed in the lattice, as the next result states.

\begin{theorem}\label{pretopartial}
	Given a lattice $(L,\preceq)$ and a local congruence $\delta$ on $L$, the preorder $\preceq_\delta$ given in Definition \ref{Def:preorden} is a partial order if and only if every $\delta$-cycle in $L$ is closed.
\end{theorem}
\begin{proof}
Let us assume that $\delta$ is a local congruence on $L$ and that every $\delta$-cycle in $L$ is closed and let us prove that $\preceq_\delta$ is a partial order.

	 The reflexivity of $\preceq_\delta$ holds in a direct way. 
	 
	 Now, we prove the transitivity. If $[x]_\delta \preceq_\delta [y]_\delta$ and $[y]_\delta \preceq_\delta [z]_\delta$ for $[x]_\delta, [y]_\delta, [z]_\delta \in L/\delta$, then there exist two $\delta$-sequences $(x', y_1)_\delta = (x', p_1, \dots, p_{n}, y_1)$ and $(y_2, z')_\delta = (y_2, q_1, \dots, q_{m}, z')$, with $x'\in [x]_\delta$, $y_1, y_2\in [y]_\delta$ and $z'\in [z]_\delta$. Hence, there exists a $\delta$-sequence $(x', z')_\delta = (x', p_1, \dots, p_n, y_1, y_2, q_1, \dots, q_m, z')$ satisfying the conditions of Definition \ref{Def:preorden}. Thus, $[x]_\delta \preceq_\delta [z]_\delta$ and the relation $\preceq_\delta$ is transitive.
	 
	 In order to prove that $\preceq_\delta$ is antisymmetric, we assume that $[x]_\delta\preceq_\delta[y]_\delta$ and $[y]_\delta\preceq_\delta[x]_\delta$, for some $x, y \in L$. Then there exist $x'\in[x]_\delta$, $y'\in[y]_\delta$, a $\delta$-sequence $(x', y')_\delta = (x', p_1, \dots, p_{n}, y')$ and a $\delta$-sequence $(y', x')_\delta = (y', q_1, \dots, q_m, x')$. Clearly, $(x',p_1,\dots,y',q_1,\dots, x')$ is a $\delta$-cycle and since every $\delta$-cycle is closed, we obtain $[x]_\delta = [y]_\delta$. Hence $\preceq_\delta$ is antisymmetric and thus a partial order.
	 
	 Now, suppose that $\preceq_\delta$ is a partial order, if $(p_0, \dots, p_n, p_0)$ is a $\delta$-cycle of $L$ (as it is showed in Figure \ref{Ex:dcycle}) then we have that
	 $$ [p_0]_\delta \circledast_1 [p_1]_\delta \circledast_2 [p_2]_\delta \circledast_3 [p_3]_\delta \circledast_4 \dots \circledast_{n-2}[p_{n-2}]_\delta \circledast_{n-1} [p_{n-1}]_\delta \circledast_n [p_n]_\delta \circledast_0 [p_0]_\delta $$
where $\circledast_i \in \{=, \preceq_\delta\}$ for all $i\in\{0, \dots, n\}$. Since the chain begins and ends with the same element, and $\preceq_\delta$ is a partial order,
we obtain that the $\delta$-cycle is closed. \qed
\end{proof}

As a consequence, under the assumption of the introduced  necessary and sufficient condition, this result allows to  order the convex sublattices (classes) obtained after the attribute reduction, which provide a hierarchization among the obtained concepts. The following example shows that  this hierarchization  does not form a complete lattice.

\begin{example}\label{ex:nolattice}
Let $(L, \preceq)$ be the  lattice given in the left side of Figure~\ref{ex:contraej} which is isomorphic to a concept lattice obtained from a formal context and $\delta$ the local congruence shown in the middle of Figure~\ref{ex:contraej}. In this case, the considered local congruence makes that the relation $\preceq_\delta$ be a partial order.

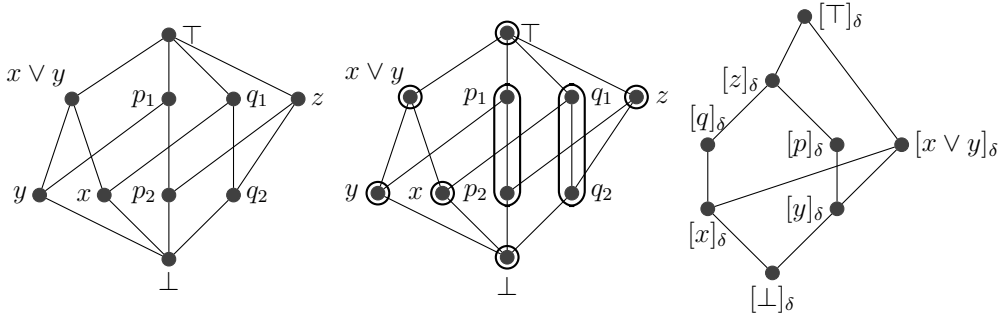
\begin{figure}[h!]
		\minipagethree{\tikzstyle{place}=[circle,draw=black!75,fill=black!75]
			\begin{tikzpicture}[inner sep=0.75mm,scale=0.85, every node/.style={scale=0.85}]	
			\node at (0,0) (b) [place, label=below:$\bot$] {};  			
			\node at (-1,1) (x) [place, label=left:$x$] {};
			\node at (-2,1) (y) [place, label=left:$y$] {};
			\node at (0,1) (p2) [place, label=left:$p_2$] {};
			\node at (1,1) (q2) [place, label=right:$q_2$] {};
			\node at (-1.5,2.5) (m) [place, label=95:$x\vee y$] {};
			\node at (0,2.5) (p1) [place, label=left:$p_1$] {};
			\node at (1,2.5) (q1) [place, label=right:$q_1$] {};
			\node at (2,2.5) (z) [place, label=right:$z$] {};
			\node at (0,3.5) (t) [place, label=right:$\top$] {};
			
			\draw [-] (b) -- (y)--(m)--(t);
			\draw [-] (b) -- (x)--(m);
			\draw [-] (b) -- (p2)--(p1)--(t);
			\draw [-] (b) -- (q2)--(q1)--(t);
			\draw [-] (y) -- (p1);
			\draw [-] (x) -- (q1);
			\draw [-] (p2) -- (z)--(t);
			\draw [-] (q2) -- (z);
			\end{tikzpicture}
			}{
			\tikzstyle{place}=[circle,draw=black!75,fill=black!75]
				\begin{tikzpicture}[inner sep=0.75mm,scale=0.85, every node/.style={scale=0.85}]
				\node at (0,0) (b) [place, label={[label distance=0.1cm]below:$\bot$}] {};  			
				\node at (-1,1) (x) [place, label={[label distance=0.1cm]left:$x$}] {};
				\node at (-2,1) (y) [place, label={[label distance=0.1cm]left:$y$}] {};
				\node at (0,1) (p2) [place, label={[label distance=0.1cm]left:$p_2$}] {};
				\node at (1,1) (q2) [place, label={[label distance=0.1cm]right:$q_2$}] {};
				\node at (-1.5,2.5) (m) [place, label=95:$x\vee y$] {};
				\node at (0,2.5) (p1) [place, label={[label distance=0.1cm]left:$p_1$}] {};
				\node at (1,2.5) (q1) [place, label={[label distance=0.1cm]right:$q_1$}] {};
				\node at (2,2.5) (z) [place, label={[label distance=0.1cm]right:$z$}] {};
				\node at (0,3.5) (t) [place, label=right:$\top$] {};
				
				\draw [-] (b) -- (y)--(m)--(t);
				\draw [-] (b) -- (x)--(m);
				\draw [-] (b) -- (p2)--(p1)--(t);
				\draw [-] (b) -- (q2)--(q1)--(t);
				\draw [-] (y) -- (p1);
				\draw [-] (x) -- (q1);
				\draw [-] (p2) -- (z)--(t);
				\draw [-] (q2) -- (z);
				\draw[thick] (b) circle (5pt);
				\draw[thick] (x) circle (5pt);	
				\draw[thick] (y) circle (5pt);	
				\draw[thick] (m) circle (5pt);	
				\draw[thick] (z) circle (5pt);	
				\draw[thick] (t) circle (5pt);			
				\draw[rounded corners=0.2cm,thick] (-0.2,0.8) rectangle (0.2,2.7);
				\draw[rounded corners=0.2cm,thick] (0.8,0.8) rectangle (1.2,2.7);
				\end{tikzpicture}
			}{
			\tikzstyle{place}=[circle,draw=black!75,fill=black!75]
			\begin{tikzpicture}[inner sep=0.75mm,scale=0.85, every node/.style={scale=0.85}]	
			\node at (0,0) (b) [place, label=below:${[\bot]_\delta}$] {};  			
			\node at (-1,1) (x) [place, label=below:${[x]_\delta}$] {};
			\node at (1,1) (y) [place, label=left:${[y]_\delta}$] {};
			\node at (1,2) (p) [place, label=left:${[p]_\delta}$] {};
			\node at (-1,2) (q) [place, label=above:${[q]_\delta}$] {};
			\node at (2,2) (m) [place, label=right:${[x\vee y]_\delta}$] {};
			\node at (0,3) (z) [place, label=left:${[z]_\delta}$] {};
			\node at (0.5,4) (t) [place, label=right:${[\top]_\delta}$] {};
			
			\draw [-] (b) -- (y)--(p)--(z)--(t);
			\draw [-] (b) -- (x)--(q)--(z);
			\draw [-] (y) -- (m)--(t);
			\draw [-] (x) -- (m);
			\end{tikzpicture}
		
			}
			\caption{The lattice $(L,\preceq)$ (left), the local congruence $\delta$ on $L$ (middle) and its corresponding quotient set $(L/\delta, \preceq_\delta)$ with the ordering relation $\preceq_\delta$ (right).} \label{ex:contraej}
	\end{figure}

However, the quotient set $L/\delta$ ordered with $\preceq_\delta$ does not form a lattice, as it is shown in the right side of Figure~\ref{ex:contraej}, because the equivalence classes $[x]_\delta$ and $[y]_\delta$ have not got a supremum, that is, their least upper bound does not exist.    \qed
\end{example}

Therefore,  an ordering can be defined on the classes of a local congruence, when every $\delta$-cycle in $L$ is closed, which could not provide a complete lattice, but it is enough to produce a hierarchization among the computed reduced concepts. In order to ensure that a local congruence can always be computed, such as every $\delta$-cycle is closed, more properties of local congruences need to be studied. Specifically, it is important to analyze the relationships among these new congruences.

\section{Algebraic structure of local congruences on a lattice}\label{algest:wc}
	In this section, we study the algebraic structure of the set of all local congruences defined on a lattice. 
First of all, we show that local congruences can be ordered by using the  definition of inclusion of  equivalence relations, which is recalled 
next.

\begin{definition}\label{weakinclusion}
	Let $\rho_1$ and $\rho_2$ be two equivalence relations on a lattice $(L,\preceq)$. We say that the equivalence relation $\rho_1$ is included in $\rho_2$,  denoted as $\rho_1 \sqsubseteq \rho_2$, if for every equivalence class $[x]_{\rho_1} \in L/ \rho_1$ there exists an equivalence class $[y]_{\rho_2} \in L /\rho_2$ such that $[x]_{\rho_1}\subseteq [y]_{\rho_2}$.
	
	We say that two equivalence relations, $\rho_1$ and $\rho_2$, are incomparable if $\rho_1 \not\sqsubseteq \rho_2$ and $\rho_2 \not\sqsubseteq \rho_1$.
\end{definition}

	From now on,  the set of all local congruences on $L$ ordered by the inclusion $\sqsubseteq$ will be denoted as $(\text{LCon}~ L, \sqsubseteq)$. 
	{First of all,  we will show that the set $(\text{LCon}~ L, \sqsubseteq)$ is a complete lattice, proving that $(\text{LCon}~ L, \sqsubseteq)$ is a topped $\sqcap$-structure with  a maximum element. In addition, the maximum and the minimun  of the complete lattice $(\text{LCon}~L,\sqsubseteq)$ are characterized.}
	
	{
	\begin{theorem}\label{inter:estruct}
	Given a lattice $(L,\preceq)$, the set $(\text{LCon}~L, \sqsubseteq)$ is a complete lattice. 
	Moreover, the least and greatest element are given by $\delta_\bot= \{(a,a) \mid a\in L\}$ and $\delta_\top=\{(a,b) \mid a,b\in L \}$, respectively. \end{theorem}
}

	\begin{proof}
		{
		Let us assume that $(L, \preceq)$ is a lattice and  $(\text{LCon}~L, \sqsubseteq)$ is the set of all local congruences. {First of all}, we need to prove that $(\text{LCon}~L, \sqsubseteq)$ is a topped $\cap$-structure. Therefore, we consider a non-empty family of local congruence, that is, $\{\delta_i\}_{i\in I} \subseteq \text{LCon}~L$ where $I$ is a index set.
		
		It is well known that the intersection of equivalence relations is an equivalence relation. Hence, $\bigcap_{i\in I} \delta_i$ is indeed an equivalence relation. Now, we prove that each equivalence class of the intersection is a convex sublattice. Let us consider an equivalence class $Z\in L/(\bigcap_{i\in I}\delta_i)$, hence there exist {a family of} equivalence classes ${\{X_i \in L/\delta_i\mid i\in I\}}$ such that $Z = \bigcap_{i\in I} X_i$. If we consider $a, b\in Z$, then we have that $a,b\in X_i$ for all $i\in I$ and, since each $X_i$ is a convex sublattice of $L$, we have that $a\wedge b, a\vee b\in X_i$ for all $i \in I$. Therefore, $a\wedge b, a\vee b\in \bigcap_{i\in I}X_i = Z$, that is, $Z$ is a sublattice of $L$. In addition, if $a\preceq b$ and we consider $c\in L$ such that $a\preceq c \preceq b$, then we have that $c\in X_i$ for all $i\in I$ since each $X_i$ is convex. Therefore, $c\in \bigcap_{i\in I} = Z$, that is, $Z$ is also convex. Thus, $\bigcap_{i\in I} \delta_i \in \text{LCon}~L$, i.e., $\text{LCon}~L$ is a $\cap$-structure.
		
		Now, we need to prove that $\text{LCon}~L$ has a maximum element. It is clear that the equivalence relation on $L$ that relates all elements of $L$, that is, $\{(a,b) \mid a,b\in L \} = L\times L$, has  convex sublattices of $L$ as equivalence classes, hence $\{(a,b) \mid a,b\in L \} = L\times L \in \text{LCon}~L$ and moreover, we cannot find another local congruence that contains it. Therefore, $\{(a,b) \mid a,b\in L \} = L\times L$ is the greatest local congruence and we denote it as $\delta_\top$. Thus, the set $(\text{LCon}~L, \sqsubseteq)$ is a complete lattice.
		
		In addition, it is clear that the least local congruence is the equivalence relation on $L$ that {only} relates each elements of $L$ to itself, that is, $\delta_\bot= \{(a,a) \mid a\in L\}$.\qed
		
	}
\end{proof}

	Next definition shows the notion of principal local congruence, which is the least local congruence that can be defined from two given elements of a lattice. 	
	\begin{definition}\label{principalcongruence}
		Given a pair of elements $(a, b)\in L\times L$, the \emph{principal local congruence generated by} $(a,b)$, denoted as $\delta_{(a,b)}$, is the least local congruence that contains the elements $a$ and $b$ in the same equivalence class{, that is 
$$
\delta_{(a,b)} = \bigwedge\{\delta\in \text{LCon}~L \mid (a,b)\in \delta\}
$$
		}
	\end{definition}

{
Note that, for every  pair of elements $(a, b)\in L\times L$, the principal local congruence $\delta_{(a,b)}$ always exists since the set $(\text{LCon}~L, \sqsubseteq)$ is a complete lattice.
}

{Finally, the last theorem generalizes the characterization of congruences in terms of principal congruences (recalled in Lemma~\ref{Ch:princongruences}) for local congruences, considering an arbitrary equivalence relation.} 

\begin{theorem}\label{th:leastlc}
	Given a lattice $(L, \preceq)$ and an equivalence relation $\rho$, the least local congruence containing $\rho$ is
	$$\delta_{\rho} = \bigvee \{\delta_{(a,b)} \mid (a,b)\in\rho \}$$
\end{theorem}
\begin{proof}
	Let us assume that $\delta_{\rho}$ is the least local congruence containing an equivalence relation $\rho$ and let us prove that $\delta_\rho$ is the least upper bound of the set $S=\{\delta_{(a,b)} \mid (a,b)\in\rho \}$. Due to $\rho\sqsubseteq \delta_{\rho}$, it is clear that $S \subseteq \{\delta_{(c,d)} \mid (c,d)\in\delta_\rho \}$ and, by Proposition~\ref{wc:pwc}, we have that $\delta_\rho = \bigvee\{\delta_{(c,d)} \mid (c,d)\in\delta_\rho \}$. Hence $\delta_\rho$ is an upper bound for $S$. Now, let us  assume that $\delta_\rho '$ is an upper bound for $S$, which means that for all $(a,b)\in \rho$ then $\delta_{(a,b)}\sqsubseteq \delta_\rho'$.
Therefore, by the supremum property we have that
	$$
	\delta_{\rho} = \bigvee \{\delta_{(a,b)} \mid (a,b)\in\rho \}\sqsubseteq \delta_\rho'
	$$
which finishes the proof.
\qed
\end{proof}

{In particular, the previous result is also satisfied when we consider a local congruence instead of an arbitrary equivalence relation.}
\begin{corollary}\label{wc:pwc}
	Let $\left( L, \preceq \right)$ be a lattice and let $\delta$  a local congruence of $(\text{LCon}~L, \sqsubseteq)$. Then
	$$\delta = \bigvee\{\delta_{(a,b)} \mid (a,b)\in\delta\}.$$
\end{corollary}
{
\begin{proof}
	Straightforwardly from Theorem~\ref{th:leastlc}, considering a local congruence $\delta$ as the equivalence relation $\rho$.\qed
\end{proof}
}

Note that this result will be very important in the reduction procedure in order to obtain 
 a local congruence $\delta$ satisfying that $(L/\delta,\preceq_\delta)$ is a partial ordered set (poset), as we will show in the next section.

\section{Reduction mechanism of concept lattices}\label{sec:reducmec}

This section will introduce an attribute reduction mechanism focused on grouping concepts in convex sublattices, having a hierarchy in form of a poset, which is equivalent by Theorem~\ref{pretopartial} to computing a local congruence with all $\delta$-cycle in $L$  being closed.
 In order to fulfill this last requirement we will use the following procedure from an arbitrary local congruence. Given a lattice $(L,\preceq)$ and a local congruence $\delta$ on $L$, if every $\delta$-cycle in $L$ is closed, then we already have that $(L/\delta,\preceq_\delta)$ is a poset. 

Otherwise, we can define an equivalence relation $\rho$ on $L/\delta$ as 
\begin{equation}\label{eq:rho}
  \rho_{\delta} = \{([x]_\delta, [y]_\delta)\in L/\delta \times L/\delta \mid [x]_\delta \preceq_\delta [y]_\delta \text{ and } [y]_\delta \preceq_\delta [x]_\delta \}
 \end{equation}
  If there are two different equivalence classes $[x]_\delta$, $[y]_\delta$ such that $[x]_\delta \preceq_\delta [y]_\delta$ and $[y]_\delta \preceq_\delta [x]_\delta$, this means that there is a $\delta$-cycle, $(x',x')_\delta$ or $(y',y')_\delta$ for some $x'\in[x]_\delta$, $y'\in [y]_\delta$. Therefore, the equivalence relation $\rho_{\delta}$ groups all the equivalence classes that contain elements in the $\delta$-cycle in a unique equivalence class providing a new partition of $L$.

However, the equivalence relation $\rho_{\delta}$ may not be a local congruence. Since clearly $\delta \sqsubseteq \rho_{\delta}$, by Theorem~\ref{th:leastlc}, we can find the least local congruence $\bar{\delta}$ that contains the  equivalence relation $\rho_{\delta}$, that is, $\delta \sqsubseteq \rho_{\delta} \sqsubseteq \bar{\delta}$. 

Hence, every $\bar{\delta}$-cycle in $L$ is closed and, by Theorem~\ref{pretopartial},  $\preceq_{\bar{\delta}}$ is a partial order on the quotient set $L/{\bar{\delta}}$. 

This procedure to ensure the ordering between the classes will be incorporate in the

procedure to reduce concept lattices by  local congruences, which is summarized in the following steps:

\begin{algorithm}

	\caption{Reducing concept lattices by  local congruences}
	\label{alg:1}
	
	\SetKwInOut{Input}{input}\SetKwInOut{Output}{output}
	
	\Input{$\mathcal C(A,B, R)$, $D\subseteq A$}
	\Output{$\delta$}
	\BlankLine
	Obtain the relation $\rho_D$ associated with the attribute reduction given by $D$\;
	Compute  the {least} local congruence $\delta_D$ containing $\rho_D$\;

	\If   {every $\delta_D$-cycle is closed,}{$\delta= \delta_D$}
	\Else{ Compute $\rho_{\delta_D} $ by Equation~\eqref{eq:rho}\;
	 \If {$\rho$ is a  local congruence}{$\delta=\rho_{\delta_D}$}
	 \Else{Obtain  the least local congruence ${\delta_\rho}$ such that $\delta_D \sqsubseteq \rho_{\delta_D} \sqsubseteq \delta_\rho$\;
$\delta= \delta_\rho$}}
\Return{$\delta$}
	\BlankLine
\end{algorithm}

Notice that, the set $D$  in Algorithm~\ref{alg:1} can be given from any reduction mechanism. For example,  it can be a rough set reduct~\cite{bmrd:RSTFCA:c,bmrd:FCARST:f}. Moreover, observe that the relation $\rho_D$ was defined in the classical case in Proposition~\ref{prop:clase2} and in the fuzzy case in Proposition~\ref{prop:partition}.

{
{This previous mechanism provides the desired reduction, as the following result shows.} 
\begin{proposition}\label{procedimiento}
	Given a concept lattice $\mathcal{C}(A, B, R)$ and a subset of attributes $D\subseteq A$, then Algorithm~\ref{alg:1} provides the least local congruence $\delta$ containing the induced {relation} $\rho_D$ and $(\mathcal{C}(A, B, R)/\delta, \preceq_{\delta})$ is a poset.
\end{proposition}
\begin{proof}
	Let us assume that we have a concept lattice $\mathcal{C}(A,B,R)$ and a partition of $\mathcal{C}(A, B, R)$ induced by an attribute reduction provided by $D\subseteq A$.  

The starting point of the procedure (Line 1) is the computation of the equivalence relation  associated with the attribute reduction given by the subset $D$, which is denoted as $\rho_D$. 

In Line 2, by   Theorem~\ref{th:leastlc}, we obtain the least local congruence containing $\rho_D$, which is denoted as $\delta_D$. Hence, in particular, $\rho_D \sqsubseteq \delta_D$.
		From this local congruence, the relation $\preceq_{\delta_D}$ defined as in Definition~\ref{Def:preorden} is a preorder.
	By  Theorem~\ref{pretopartial},  if  every $\delta_D$-cycle is closed (checked in Line 3), then  $(\mathcal{C}(A, B, R)/\delta, \preceq_{\delta})$ is a poset, that is, the required relation $\delta$ is $\delta_D$ (Line 4).

	Otherwise,  $\preceq_{\delta_D}$ is only a preorder and we consider the new equivalence relation 
		$\rho_{\delta_D}$ defined in Equation~\ref{eq:rho}. 
As a consequence, we have that $\delta_D \sqsubseteq \rho_{\delta_D}$.
If $\rho_{\delta_D}$ is a local congruence, by the definition of $\rho_{\delta_D}$, we have that every $\rho_{\delta_D}$-cycle is closed and, according to  Theorem~\ref{pretopartial}, $\preceq_{\rho_{\delta_D}}$ is a partial order on $\mathcal{C}(A, B, R)/\rho_{\delta_D}$. In this case,  $\delta=\rho_{\delta_D}$  is the least local congruence we are interested in (Lines 6-8).
 Otherwise, from   Theorem~\ref{th:leastlc}, in Line 10 we obtain the least local congruence $\delta_\rho$ containing to $\rho_{\delta_D}$, such that $\rho_{\delta_D}\sqsubseteq\delta_\rho$. Therefore, we have that every $\delta_\rho$-cycle is closed by the definition of $\rho_{\delta_D}$, and  by Theorem~\ref{pretopartial} we obtain that  $\preceq_{\delta_\rho}$ is a partial order on $\mathcal{C}(A, B, R)/\delta_\rho$. Thus,     $\delta=\delta_\rho$ is the least local congruence we are looking for.

Consequently, from the procedure we obtain that $(\mathcal{C}(A, B, R)/\delta, \preceq_{\delta})$ is a poset where $\delta$ is the least local congruence containing $\rho_D$.\qed
\end{proof}
}
In the next example we will show the procedure described above.

\begin{example}\label{Ex:procedure}

Let us consider {a context $(A, B, R)$ and a subset of attributes $D\subseteq A$ such that after the reduction process we obtain the induced partition of the concept lattice displayed in the left side of Figure~\ref{wc:procedure}. Thus, we consider the local congruence $\delta_D$ displayed in the right side of Figure~\ref{wc:procedure}, it is easy to check that $\delta_D$ is indeed a local congruence and the least one containing the induced partition.  }
		\begin{figure}[h!]		
		
		\minipagetwo{
			\tikzstyle{place}=[circle,draw=black!75,fill=black!75]
			\begin{tikzpicture}[inner sep=0.75mm,scale=0.9, every node/.style={scale=0.9}]		
			\node at (0,-1) (bot) [place, label={[label distance=0.1cm]below:$\bot$}] {};
			\node at (-2,0) (p0) [place, label={[label distance=0.1cm]left:$p_0$}] {};  			
			\node at (0,0) (p1) [place, label={[label distance=0.1cm]left:$p_1$}] {};
			\node at (2,0) (p2) [place, label={[label distance=0.1cm]right:$p_2$}] {};
			\node at (-2,2) (p9) [place, label=left:$p_9$] {};  			
			\node at (0,2) (p10) [place, label={[label distance=0.1cm]left:$p_{10}$}] {};
			\node at (2,2) (p11) [place, label={[label distance=0.1cm]right:$p_{11}$}] {};
			\node at (0,3) (p12) [place, label={[label distance=0.1cm]left:$p_{12}$}] {};
			\node at (0,4) (p13) [place, label={[label distance=0.1cm]right:$p_{13}$}] {};
			\node at (-0.75,4.5) (p14) [place, label={[label distance=0.1cm]left:$p_{14}$}] {};
			\node at (-0.75,5.5) (p16) [place, label={[label distance=0.1cm]left:$p_{16}$}] {};
			\node at (0.75,5) (p15) [place, label={[label distance=0.1cm]right:$p_{15}$}] {};
			\node at (0,6) (top) [place, label={[label distance=0.1cm]above:$\top$}] {};
			\node at (-1.5,1) (p4) [place, label=175:$p_4$] {};
			\node at (-2.5,1) (p3) [place, label=left:$p_3$] {};
			\node at (-0.5,1) (p5) [place, label=left:$p_5$] {};
			\node at (0.5,1) (p6) [place, label=175:$p_6$] {};
			\node at (1.5,1) (p7) [place, label=left:$p_7$] {};
			\node at (2.5,1) (p8) [place, label={[label distance=0.1cm]right:$p_8$}] {};
			
			\draw[-] (bot)--(p0)--(p4)--(p9)--(p12)--(p13)--(p14)--(p16)--(top)--(p15)--(p13);
			\draw[-] (p0)--(p3)--(p9);
			\draw[-] (p0)--(p5)--(p10)--(p6)--(p1)--(bot)--(p2)--(p7)--(p11)--(p8)--(p2)--(p3);
			\draw[-] (p10)--(p12)--(p11);
			\draw[-] (p5)--(p1)--(p7);
		
		\draw[ rounded corners=0.2cm,thick, densely dotted] (-2.7,0.85)--++(0.7,1.5)--++(0.75,-1.5)--cycle;	
		\draw[ rounded corners=0.2cm,thick, densely dotted] (-0.7,0.85)--++(0.7,1.5)--++(0.75,-1.5)--cycle;
		\draw[ rounded corners=0.2cm,thick, densely dotted] (1.3,0.85)--++(0.7,1.5)--++(0.75,-1.5)--cycle;
		\draw[thick, densely dotted] (p12) circle (6pt);
		\draw[thick, densely dotted] (p1) circle (6pt);
		\draw[thick, densely dotted] (p2) circle (6pt);
		\draw[thick, densely dotted] (p0) circle (6pt);
		\draw[thick, densely dotted] (bot) circle (6pt);
		\draw[ rounded corners=0.2cm,thick, densely dotted] (-0.95,4.3) rectangle (-0.55,5.7);
		\draw[rotate around={37.5:(0.,5.3)}, rounded corners=0.2cm,thick, densely dotted] (0.2,4.4) rectangle (0.65,6.1);
		\draw[thick, densely dotted] (p13) circle (6pt);
			\end{tikzpicture}
		}{ 
		\tikzstyle{place}=[circle,draw=black!75,fill=black!75]
		\begin{tikzpicture}[inner sep=0.75mm,scale=0.9, every node/.style={scale=0.9}]		
		\node at (0,-1) (bot) [place, label={[label distance=0.1cm]below:$\bot$}] {};
		\node at (-2,0) (p0) [place, label={[label distance=0.1cm]left:$p_0$}] {};  			
		\node at (0,0) (p1) [place, label={[label distance=0.1cm]left:$p_1$}] {};
		\node at (2,0) (p2) [place, label={[label distance=0.1cm]right:$p_2$}] {};
		\node at (-2,2) (p9) [place, label=left:$p_9$] {};  			
		\node at (0,2) (p10) [place, label={[label distance=0.1cm]left:$p_{10}$}] {};
		\node at (2,2) (p11) [place, label={[label distance=0.1cm]right:$p_{11}$}] {};
		\node at (0,3) (p12) [place, label={[label distance=0.1cm]left:$p_{12}$}] {};
		\node at (0,4) (p13) [place, label={[label distance=0.1cm]right:$p_{13}$}] {};
		\node at (-0.75,4.5) (p14) [place, label={[label distance=0.1cm]left:$p_{14}$}] {};
		\node at (-0.75,5.5) (p16) [place, label={[label distance=0.1cm]left:$p_{16}$}] {};
		\node at (0.75,5) (p15) [place, label={[label distance=0.1cm]right:$p_{15}$}] {};
		\node at (0,6) (top) [place, label={[label distance=0.1cm]above:$\top$}] {};
		\node at (-1.5,1) (p4) [place, label=175:$p_4$] {};
		\node at (-2.5,1) (p3) [place, label=left:$p_3$] {};
		\node at (-0.5,1) (p5) [place, label=left:$p_5$] {};
		\node at (0.5,1) (p6) [place, label=175:$p_6$] {};
		\node at (1.5,1) (p7) [place, label=left:$p_7$] {};
		\node at (2.5,1) (p8) [place, label={[label distance=0.1cm]right:$p_8$}] {};
		
		\draw[-] (bot)--(p0)--(p4)--(p9)--(p12)--(p13)--(p14)--(p16)--(top)--(p15)--(p13);
		\draw[-] (p0)--(p3)--(p9);
		\draw[-] (p0)--(p5)--(p10)--(p6)--(p1)--(bot)--(p2)--(p7)--(p11)--(p8)--(p2)--(p3);
		\draw[-] (p10)--(p12)--(p11);
		\draw[-] (p5)--(p1)--(p7);
		
		\draw[ rounded corners=0.2cm,thick] (-2.7,0.85)--++(0.7,1.5)--++(0.7,-1.4)--++(-0.7,-1.3)--cycle;	
		\draw[ rounded corners=0.2cm,thick] (-0.7,0.85)--++(0.7,1.5)--++(0.7,-1.4)--++(-0.7,-1.3)--cycle;
		\draw[ rounded corners=0.2cm,thick] (1.3,0.85)--++(0.7,1.5)--++(0.7,-1.4)--++(-0.7,-1.3)--cycle;
		\draw[thick] (p12) circle (6pt);
		\draw[thick] (bot) circle (6pt);
		\draw[ rounded corners=0.2cm,thick] (-0.95,4.3) rectangle (-0.55,5.7);
		\draw[rotate around={37.5:(0.,5.3)}, rounded corners=0.2cm,thick] (0.2,4.4) rectangle (0.65,6.1);
		\draw[thick] (p13) circle (6pt);
		\end{tikzpicture}
		}
		
		\caption{Partition induced $\rho_D$ (left) and local congruence $\delta_D$ (right) of Example~\ref{Ex:procedure}.}
		\label{wc:procedure}	
	\end{figure}
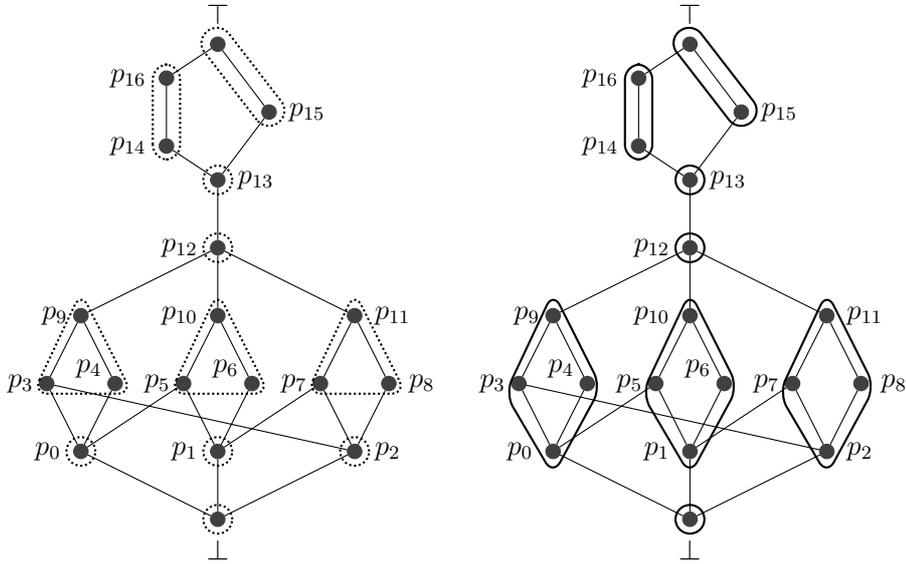

Considering the relation $\preceq_{\delta_D}$  given as in Definition~\ref{Def:preorden}, we can note that the $\delta_D$-sequence, $(p_0,p_0)_{\delta_D} = (p_0, p_5, p_1, p_7, p_2, p_3, p_0)$, is in fact a $\delta_D$-cycle in the lattice and it is not closed. 
Therefore, we define the equivalence relation $\rho =  \{([x]_{\delta_D}, [y]_{\delta_D})\in L/{\delta_D} \times L/{\delta_D} \mid [x]_{\delta_D} \preceq_{\delta_D} [y]_{\delta_D} \text{ and } [y]_{\delta_D} \preceq_{\delta_D} [x]_{\delta_D} \}$. The new partition of $L$  provided by the equivalence relation $\rho$ is shown in the left side of Figure~\ref{equivrelwc}. We can observe that the equivalence relation $\rho$ groups the classes of $L/{\delta_D}$ that contain elements in  the ${\delta_D}$-cycle into a single equivalence class. Moreover,  it is also easy to observe that ${\delta_D} \sqsubseteq \rho$.
	
Now, we have to verify if the equivalence relation $\rho$ is a local congruence, but we can observe that it is not, since the equivalence class that contains the $\delta_D$-cycle is not a (convex) sublattice of $L$. Thus, we must find the least local congruence ${\delta_\rho}$ that contains the equivalence relation $\rho$.

		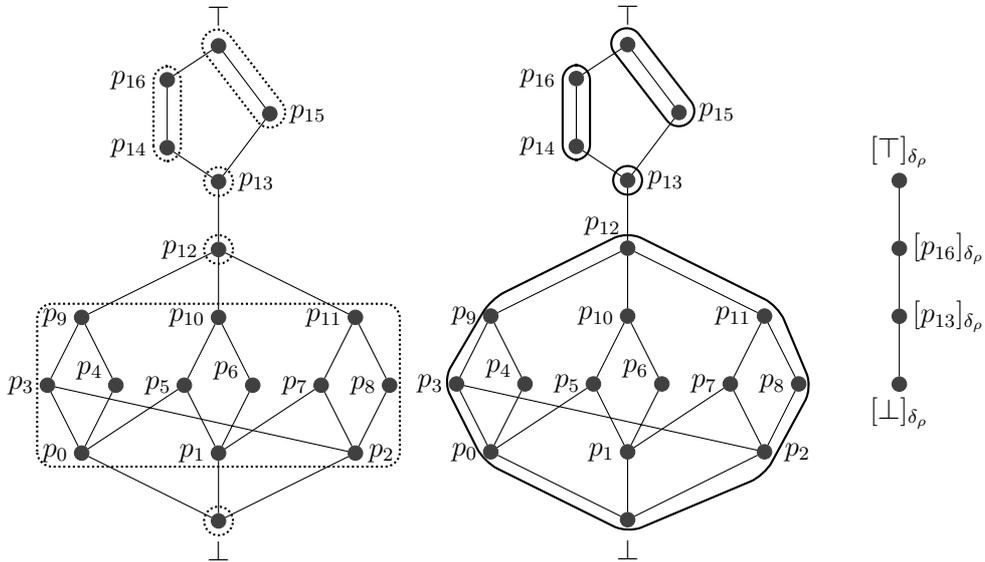
\begin{figure}[h!]
		\begin{minipage}{0.38\textwidth}
		\begin{center}		
		\tikzstyle{place}=[circle,draw=black!75,fill=black!75]
		\begin{tikzpicture}[inner sep=0.75mm,scale=0.9, every node/.style={scale=0.9}]		
			\node at (0,-1) (bot) [place, label={[label distance=0.1cm]below:$\bot$}] {};
			\node at (-2,0) (p0) [place, label={left:$p_0$}] {};  			
			\node at (0,0) (p1) [place, label={left:$p_1$}] {};
			\node at (2,0) (p2) [place, label={right:$p_2$}] {};
			\node at (-2,2) (p9) [place, label=left:$p_9$] {};  			
			\node at (0,2) (p10) [place, label={left:$p_{10}$}] {};
			\node at (2,2) (p11) [place, label={left:$p_{11}$}] {};
			\node at (0,3) (p12) [place, label={[label distance=0.1cm]left:$p_{12}$}] {};
			\node at (0,4) (p13) [place, label={[label distance=0.1cm]right:$p_{13}$}] {};
			\node at (-0.75,4.5) (p14) [place, label={[label distance=0.1cm]left:$p_{14}$}] {};
			\node at (-0.75,5.5) (p16) [place, label={[label distance=0.1cm]left:$p_{16}$}] {};
			\node at (0.75,5) (p15) [place, label={[label distance=0.1cm]right:$p_{15}$}] {};
			\node at (0,6) (top) [place, label={[label distance=0.1cm]above:$\top$}] {};
			\node at (-1.5,1) (p4) [place, label=175:$p_4$] {};
			\node at (-2.5,1) (p3) [place, label=left:$p_3$] {};
			\node at (-0.5,1) (p5) [place, label=left:$p_5$] {};
			\node at (0.5,1) (p6) [place, label=175:$p_6$] {};
			\node at (1.5,1) (p7) [place, label=left:$p_7$] {};
			\node at (2.5,1) (p8) [place, label={left:$p_8$}] {};
			
			\draw[-] (bot)--(p0)--(p4)--(p9)--(p12)--(p13)--(p14)--(p16)--(top)--(p15)--(p13);
			\draw[-] (p0)--(p3)--(p9);
			\draw[-] (p0)--(p5)--(p10)--(p6)--(p1)--(bot)--(p2)--(p7)--(p11)--(p8)--(p2)--(p3);
			\draw[-] (p10)--(p12)--(p11);
			\draw[-] (p5)--(p1)--(p7);
			
			\draw[ rounded corners=0.2cm,thick, densely dotted] (-2.65,-0.2) rectangle (2.65,2.2);
			\draw[thick, densely dotted] (p12) circle (6pt);
			\draw[thick, densely dotted] (bot) circle (6pt);
			\draw[ rounded corners=0.2cm,thick, densely dotted] (-0.95,4.3) rectangle (-0.55,5.7);
			\draw[rotate around={37.5:(0.,5.3)}, rounded corners=0.2cm,thick, densely dotted] (0.2,4.4) rectangle (0.65,6.1);
			\draw[thick, densely dotted] (p13) circle (6pt);
		\end{tikzpicture}
			\end{center}
	\end{minipage}
\begin{minipage}{0.41\textwidth}
\begin{center}	
		\tikzstyle{place}=[circle,draw=black!75,fill=black!75]
		\begin{tikzpicture}[inner sep=0.75mm,scale=0.9, every node/.style={scale=0.9}]		
		\node at (0,-1) (bot) [place, label={[label distance=0.1cm]below:$\bot$}] {};
		\node at (-2,0) (p0) [place, label={left:$p_0$}] {};  			
		\node at (0,0) (p1) [place, label={left:$p_1$}] {};
		\node at (2,0) (p2) [place, label={[label distance=0.1cm]right:$p_2$}] {};
		\node at (-2,2) (p9) [place, label=left:$p_9$] {};  			
		\node at (0,2) (p10) [place, label={left:$p_{10}$}] {};
		\node at (2,2) (p11) [place, label={left:$p_{11}$}] {};
		\node at (0,3) (p12) [place, label={100:$p_{12}$}] {};
		\node at (0,4) (p13) [place, label={[label distance=0.1cm]right:$p_{13}$}] {};
		\node at (-0.75,4.5) (p14) [place, label={[label distance=0.1cm]left:$p_{14}$}] {};
		\node at (-0.75,5.5) (p16) [place, label={[label distance=0.1cm]left:$p_{16}$}] {};
		\node at (0.75,5) (p15) [place, label={[label distance=0.1cm]right:$p_{15}$}] {};
		\node at (0,6) (top) [place, label={[label distance=0.1cm]above:$\top$}] {};
		\node at (-1.5,1) (p4) [place, label=175:$p_4$] {};
		\node at (-2.5,1) (p3) [place, label=left:$p_3$] {};
		\node at (-0.5,1) (p5) [place, label=left:$p_5$] {};
		\node at (0.5,1) (p6) [place, label=175:$p_6$] {};
		\node at (1.5,1) (p7) [place, label=left:$p_7$] {};
		\node at (2.5,1) (p8) [place, label={left:$p_8$}] {};
		
		\draw[-] (bot)--(p0)--(p4)--(p9)--(p12)--(p13)--(p14)--(p16)--(top)--(p15)--(p13);
		\draw[-] (p0)--(p3)--(p9);
		\draw[-] (p0)--(p5)--(p10)--(p6)--(p1)--(bot)--(p2)--(p7)--(p11)--(p8)--(p2)--(p3);
		\draw[-] (p10)--(p12)--(p11);
		\draw[-] (p5)--(p1)--(p7);
		
		\draw[ rounded corners=0.2cm,thick] (0,-1.2)--++(-2.1,1.05)--++(-0.6,1.2)--++(0.7,1.2)--++(2,1.)--++(2.2,-1.1)--++(0.5,-1.2)--++(-0.7,-1.3)--cycle;
		\draw[ rounded corners=0.2cm,thick] (-0.95,4.3) rectangle (-0.55,5.7);
		\draw[rotate around={37.5:(0.,5.3)}, rounded corners=0.2cm,thick] (0.2,4.4) rectangle (0.65,6.1);
		\draw[thick] (p13) circle (6pt);
		\end{tikzpicture}
	\end{center}
\end{minipage}
\begin{minipage}{0.18\textwidth}
\begin{center}
		\tikzstyle{place}=[circle,draw=black!75,fill=black!75]
		\begin{tikzpicture}[inner sep=0.75mm,scale=0.9, every node/.style={scale=0.9}]		
		\node at (0,0) (0) [place, label=below:${[\bot]_{\delta_\rho}}$] {};
		\node at (0,1) (d0) [place, label=right:${[p_{13}]_{\delta_\rho}}$] {};
		\node at (0,2) (d1) [place, label=right:${[p_{16}]_{\delta_\rho}}$] {};
		\node at (0,3) (d2) [place, label=above:${[\top]_{\delta_\rho}}$] {};
		
		\draw [-] (0)--(d0)--(d1)--(d2);
		\end{tikzpicture}
	\end{center}
\end{minipage}
		\caption{The equivalence relation $\rho$ on $L/\delta$ (left), the least local congruence $\delta_\rho$ containing $\rho$ (middle) and the quotient set $L/{\delta_\rho}$ (right).}
		\label{equivrelwc}	
	\end{figure}

In this case, the least local congruence that satisfies $ \delta_D \sqsubseteq \rho \sqsubseteq \delta_\rho$ is the local congruence shown in the middle of Figure~\ref{equivrelwc}. Hence, by Theorem~\ref{pretopartial}, we have that $\preceq_{\delta_\rho}$ is a partial order on $L/{\delta_\rho}$  and the elements of the corresponding quotient set can be ranked.  The ordered set $(L/{\delta_\rho}, \preceq_{\delta_\rho})$, is displayed in the right side of Figure~\ref{equivrelwc}.
It is important to note that the local congruence that we have finally obtained is not a congruence because the least congruence, containing the equivalence relation $\rho$, should include $p_{13}, p_{14}, p_{15}, p_{16}$ and $\top$ in the same class, in order to satisfy the quadrilateral-closed property. \qed

\end{example}

{
Now, we apply the proposed mechanism to reduce a concept lattice in a fuzzy formal concept framework. Specifically, the following example considers a fuzzy formal context studied  in~\cite{bmrd:FCARST:f}.

\begin{example}\label{Ex:18fss}
	 The considered framework  is $(L,L,L,\&^{*}_{G})$, where the lattice $L=\{0, 0.5, 1\}$ and $\&^{*}_{G}$ is the discretization of the G\"{o}del conjunctor defined on $L$.
	It is also considered a fuzzy context $(A,B,R,\sigma)$, composed of three objects, ${B=\{b_1, b_2, b_3\}}$, four attributes ${A=\{a_1, a_2, a_3, a_4\}}$, the relation $R$  shown in Table~\ref{tabla:fuzzyrel}, and the mapping $\sigma$ constantly $\&^{*}_{G}$. All concepts of this fuzzy context are listed in Figure~\ref{nex:fig1}, where the corresponding concept lattice is illustrated as well.
	\setlength{\tabcolsep}{0.5cm}
	\begin{table}[h]
		\begin{center}
			\begin{tabular}{l|c c c}
				\hline
				$R$ & $b_1$  & $b_2$ & $b_3$\\ \hline
				$a_1$ & 1 & 0 & 0 \\ 
				$a_2$ & 0 & 0.5 & 0 \\
				$a_3$ & 0 & 0 & 1 \\
				$a_4$ & 0 & 0.5 & 1 \\\hline
			\end{tabular}
		\end{center}
		\caption{Fuzzy relation $R$ of Example~\ref{Ex:18fss}.
		\label{tabla:fuzzyrel}
		}
	\end{table}	

In~\cite{bmrd:FCARST:f}, authors obtained four different reducts to reduce the concept lattice. In this example, we will consider one of these reducts, specifically  $D_1 =\{a_1, a_2\}$, to compute a local congruence of  the reduced concept lattice obtained from this reduct.

\begin{figure}[h!]
		\begin{minipage}{0.5\textwidth}
			\begin{center}
				\begin{tabular}{@{\extracolsep{15pt}}l @{}l@{}l@{}l@{} @{}l@{}l@{}l@{}l@{}}
					\hline
					$C_i$ & \multicolumn{3}{@{}l}{Extent} & \multicolumn{4}{@{}l}{Intent}\\ \cline{2-4}\cline{5-8}
					& $b_1$  & $b_2$ & $b_3$& $a_1$& $a_2$ & $a_3$ & $a_4$ \\ \hline
					0 & 0 & 0 & 0  & 1 & 1 & 1 & 1\\ 
					1 & 1 & 0 & 0  & 1 & 0 & 0 & 0\\
					2 & 0 & 0.5 & 0  & 0 & 1 & 0 & 1\\
					3 & 0 & 0 & 1  & 0 & 0 & 1 & 1\\
					4 & 1 & 1 & 1  & 0 & 0 & 0 & 0\\
					5 & 0 & 1 & 0  & 0 & 0.5 & 0 & 0.5\\
					6 & 0 & 0.5 & 1  & 0 & 0 & 0 & 1\\
					7 & 0 & 1 & 1  & 0 & 0 & 0 & 0.5\\ \hline
				\end{tabular}
			\end{center}
		\end{minipage}
		\begin{minipage}{0.45\textwidth}
			\begin{center}
				\tikzstyle{place}=[circle,draw=black!75,fill=white!20, text width= 9pt]
				\begin{tikzpicture}[inner sep=0.75mm,scale=1.1, every node/.style={scale=0.9}]		
				\node at (0,0) (c0) [place] {$C_0$};
				\node at (-1,3) (c1) [place] {$C_1$};
				\node at (0,1) (c2) [place] {$C_2$};
				\node at (1,1) (c3) [place] {$C_3$};
				\node at (0,4) (c4) [place] {$C_4$};
				\node at (0,2) (c5) [place] {$C_5$};
				\node at (1,2) (c6) [place] {$C_6$};
				\node at (0,3) (c7) [place] {$C_7$};
				
				\draw [-] (c0) -- (c2)-- (c5)--(c7)--(c4);
				\draw [-] (c0) -- (c1)-- (c4);
				\draw [-] (c0) -- (c3)-- (c6)--(c7);
				\draw [-] (c2) -- (c6);
				\end{tikzpicture}
			\end{center}
		\end{minipage}	
		\caption{Fuzzy concepts (left) and concept lattice (right) of the context associated with Table~\ref{tabla:fuzzyrel}.}
		\label{nex:fig1}
	\end{figure}
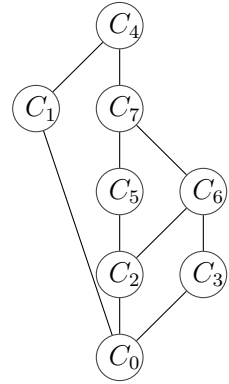
The  partition induced by $D_1$ is given in the left side of Figure~\ref{nex:fig4}, and the corresponding reduced concept lattice is depicted in its right side.
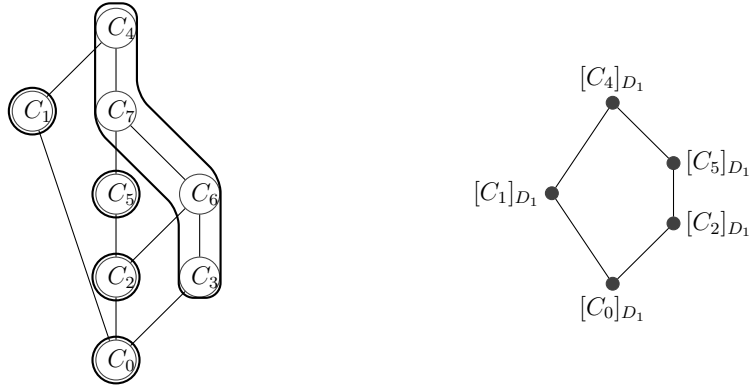
\begin{figure}[h!]
	\begin{minipage}{0.5\textwidth}
		\begin{center}
			\tikzstyle{place}=[circle,draw=black!75,fill=white!20, text width= 8pt]
			\begin{tikzpicture}[inner sep=0.75mm,scale=1.1, every node/.style={scale=0.8}]		
			\node at (0,0) (c0) [place] {$C_0$};
			\node at (-1,3) (c1) [place] {$C_1$};
			\node at (0,1) (c2) [place] {$C_2$};
			\node at (1,1) (c3) [place] {$C_3$};
			\node at (0,4) (c4) [place] {$C_4$};
			\node at (0,2) (c5) [place] {$C_5$};
			\node at (1,2) (c6) [place] {$C_6$};
			\node at (0,3) (c7) [place] {$C_7$};
			
			\draw [-] (c0) -- (c2)-- (c5)--(c7)--(c4);
			\draw [-] (c0) -- (c1)-- (c4);
			\draw [-] (c0) -- (c3)-- (c6)--(c7);
			\draw [-] (c2) -- (c6);
			\draw[thick] (c0) circle (8pt);
			\draw[thick] (c1) circle (8pt);
			\draw[thick] (c2) circle (8pt);
			\draw[thick] (c5) circle (8pt);
			\draw[rounded corners=0.2cm,thick]	(0.75,0.75)--++(0,1.05)--++(-1,1)--++(0,1.5)--++(0.5,0)--++(0,-1.1)--++(1,-1)--++(0,-1.45)--cycle;		
			\end{tikzpicture}
		\end{center}
	\end{minipage}
	\begin{minipage}{0.45\textwidth}
		\begin{center}
			\tikzstyle{place}=[circle,draw=black!75,fill=black!75]
			\begin{tikzpicture}[inner sep=0.75mm,scale=0.8, every node/.style={scale=0.8}]		
			\node at (0,0) (c0) [place, label=below:${[C_0]_{D_1}}$] {};
			\node at (-1,1.5) (c1) [place, label=left:${[C_1]_{D_1}}$] {};
			\node at (1,2) (c2) [place, label=right:${[C_5]_{D_1}}$] {};
			\node at (1,1) (c3) [place, label=right:${[C_2]_{D_1}}$] {};
			\node at (0,3) (c4) [place, label=above:${[C_4]_{D_1}}$] {};
			
			\draw [-] (c0) -- (c3)-- (c2)--(c4);
			\draw [-] (c0) -- (c1)-- (c4);
			\end{tikzpicture}
		\end{center}
	\end{minipage}	
	\caption{Partition induced by the reduction (left) and concept lattice of the reduced context (right) considering the reduct $D_1$.}
	\label{nex:fig4}
\end{figure}
 In this case, the least local congruence containing this partition is the partition itself, since each equivalence class is a convex sublattice of the original concept lattice. 
Moreover, this local congruence is not a congruence because it is not quadrilateral-closed, for example, $C_0, C_2$ and $C_3,C_6$ are opposite sides of the quadrilateral  $\langle C_0, C_2;C_3,C_6\rangle$,  the concepts $C_3$ and $C_6$ belong to one equivalence class, but  $C_0$ and $C_2$ belong to different equivalence classes.
Indeed, the least congruence containing the partition induced by the reduction of $D_1$ is the congruence with only one class containing all concepts. Thus, also in the fuzzy framework,  local congruences  offer more suitable reductions than the ones given by congruences. \qed

\end{example}
}

Therefore, the proposed reduction mechanism based on local congruences  minimizes the amount of lost information with respect to the use of congruences, clustering the concepts in convex sublattices and forming a hierarchy among them.

\section{Conclusions and future work}\label{conclusiones}

In this work, we have  introduced a weaker notion of congruence, which has been called local congruence. We have analyzed how the elements of the quotient set generated by a local congruence can be ordered.  Furthermore, we have proven that the algebraic structure of the set of local congruences is a complete lattice.  We  have also shown a characterization of local congruences in terms of its principal local congruences, as well as an extension of this characterization by considering any arbitrary equivalence relation. As a consequence, a procedure for computing the least local congruence  containing a given equivalence relation has been presented.
From this study, we have presented a new mechanism to reduce (fuzzy) concept lattices based on the notion of local congruence. Considering this reduction  mechanism, we obtain a partition of the concepts of the original concept lattice satisfying that each equivalence class has the structure of a convex sublattice of the original concept lattice. In addition, we have  shown that the consideration of local congruences to reduce concept lattices is more suitable than the consideration of congruences since a smaller amount of information is lost during the reduction process.

In the near future, more properties of the introduced procedure will be studied. For example, due to   this reduction    modifies the original partition given by the attribute reduction, it is important to analyze how it alters the formal context.  

{In addition, we are interested in studying how an optimal reduct can be selected and  the influence that this selection has on the complementary local congruence. Another important goal will be to apply this reduction procedure in real databases. Specifically, we would like to analyze the potential of the presented reduction mechanism in databases related to digital forensic analysis, in which we are leading the COST Action: DIGital FORensics:
evidence Analysis via intelligent Systems and Practices (DigForASP).}

\end{document}